\documentclass[12pt]{elsarticle}

\journal{Nucl. Instrum. Methods A}

\usepackage[margin={2cm}]{geometry}
\usepackage{rotating}
\usepackage{multirow}
\usepackage{soul}
\usepackage{xcolor}
\usepackage{chngcntr}

\usepackage{amsmath}    %for subequations
\usepackage{xspace}     %to get spaces in commands right
\usepackage{units}      %to get proper nobreak spaces for units
\usepackage{hyperref}   %clickable links in pdfs
\hypersetup{}
\usepackage[english]{babel} %bibliography
\usepackage{color}
\usepackage{multicol}
\usepackage{threeparttable}
\usepackage{refcount}
\usepackage[switch,modulo]{lineno}     %for line numbers
\newif\ifpdf
\ifx\pdfoutput\undefined
  \pdffalse
\else
  \pdfoutput=1
  \pdftrue
\fi
\ifpdf
  \usepackage{graphicx}
  \usepackage{epstopdf}
\else
  \usepackage{graphicx}
\fi
\begin{document}

\renewcommand*{\thefootnote}{\fnsymbol{footnote}}
\begin{frontmatter}
  \title{Timing Detectors with SiPM read-out for the MUSE Experiment at PSI}
  \author[Rutgers,PSI]{T.~Rostomyan}
\cortext[mycorrespondingauthor]{Dr. Tigran Rostomyan (Tigran.Rostomyan@psi.ch)}
\author[Rutgers,StonyBrook]{E.~Cline}
\author[Washington,UMich]{I.~Lavrukhin}
\author[Temple]{H.~Atac}
\author[Temple]{A.~Atencio}
\author[StonyBrook,RBRC]{J.~C.~Bernauer}
\author[Washington]{W.~J.~Briscoe}
\author[HUJI]{D.~Cohen}
\author[TAU]{E.~O.~Cohen}
\author[Washington]{C.~Collicott}
\author[PSI]{K.~Deiters}
\author[Rutgers]{S.~Dogra}
\author[Washington]{E.~Downie}
\author[UniBas]{W.~Erni}
\author[Hampton]{I.~P.~Fernando}
\author[USC]{A.~Flannery}
\author[Hampton]{T.~Gautam}
\author[UniBas]{D.~Ghosal}
\author[Rutgers]{R.~Gilman}
\author[Washington]{A.~Golossanov}
\author[Washington]{J.~Hirschman}
\author[UMich]{M.~Kim}
\author[Hampton]{M.~Kohl}
\author[UniBas]{B.~Krusche}
\author[USC]{L.~Li}
\author[Rutgers]{W.~Lin}
\author[Hampton]{A.~Liyanage}
\author[UMich]{W.~Lorenzon}
\author[MIT]{P.~Mohanmurthy}
\author[Hampton]{J.~Nazeer}
\author[HUJI]{P.~Or}
\author[Hampton]{T.~Patel}
\author[TAU]{E.~Piasetzky}
\author[TAU]{N.~Pilip}
\author[UMich]{H.~Reid}
\author[Argonne]{P.~E.~Reimer}
\author[HUJI]{G.~Ron}
\author[Washington]{E.~Rooney}
\author[TAU]{Y.~Shamai}
\author[Washington]{P.~Solazzo}
\author[USC]{S.~Strauch}
\author[TAU]{D.~Vidne}
\author[UMich]{N.~Wuerfel}

\address[Rutgers]{Department of Physics and Astronomy, Rutgers, The State University of New Jersey, Piscataway, New Jersey, 08855, USA}
\address[PSI]{Paul Scherrer Institut, Villigen, CH-5232, Switzerland}
\address[StonyBrook]{Department of Physics and Astronomy, Stony Brook University, Stony Brook, NY, 11794, USA}
\address[Washington]{Department of Physics, The George Washington University, Washington, D.C. 20052, USA}
\address[UMich]{Randall Laboratory of Physics, University of Michigan, Ann Arbor, MI 48109, USA}
\address[Temple]{Department of Physics, Temple University, Philadelphia, 19122, USA}
\address[RBRC]{Riken BNL Research Center,  Brookhaven National Laboratory, Upton, NY, 11973, USA}
\address[HUJI]{Racah Institute of Physics, The Hebrew University of Jerusalem, Jerusalem, Israel }
\address[TAU]{School of Physics and Astronomy, Tel Aviv University, Tel Aviv, 69978, Israel}
\address[UniBas]{Department of Physics, University of Basel, 4056 Basel, Switzerland}
\address[Hampton]{Physics Department, Hampton University, Hampton , VA, 23668, USA}
\address[MIT]{Laboratory for Nuclear Science, Massachusetts Institute of Technology, Cambridge, MA, 02139, USA}
\address[Argonne]{Physics Division, Argonne National Laboratory, Lemont, IL, 60439, USA}
\address[USC]{Department of Physics and Astronomy, University of South Carolina, Columbia, SC, 29208, USA}

  \begin{abstract}
The Muon Scattering Experiment at the Paul Scherrer Institut uses a mixed beam of electrons, muons, and pions, necessitating precise timing to identify the beam particles and reactions they cause.
We describe the design and performance of three timing detectors using plastic scintillator read out with silicon photomultipliers that have been built for the experiment.
The Beam Hodoscope, upstream of the scattering target, counts the beam flux and precisely times beam particles both to identify species and provide a starting time for time-of-flight measurements. 
The Beam Monitor, downstream of the scattering target, counts the unscattered beam flux, helps identify 
background in scattering events, and precisely times beam particles for time-of-flight measurements. 
The Beam Focus Monitor, mounted on the target ladder under the liquid hydrogen target inside the target vacuum chamber, is used in dedicated runs to sample the beam spot at three points near the target center, where the beam should be focused.
\end{abstract}

\end{frontmatter}

\renewcommand*{\thefootnote}{\arabic{footnote}}
\setcounter{footnote}{0}
%\begin{multicols}{2}
%\tableofcontents
%\end{multicols}
%\vspace*{0.25in}\hrule\vspace*{0.25in}

\sloppy
\begin{twocolumn}
%\linenumbers

\section{Introduction}
\label{sec:Intro}
In 2010, a Paul Scherrer Institut (PSI) experiment \cite{Pohl:2010zza} reported that the proton charge radius determined 
from  muonic hydrogen level transitions was $0.84184 \pm 0.00067$~fm, about 
5$\sigma$ off from the nearly order-of-magnitude less precise, non-muonic
measurements \citep{CODATA2014}. 
This ``proton radius puzzle'' was confirmed in 2013 by a
second measurement of muonic hydrogen \cite{Antognini:2013} that determined the radius
to be $0.84087 \pm 0.00039$~fm.
Contemporaneous electronic results of 0.879 $\pm$ 0.008 fm \cite{Bernauer:2010wm} and 0.875 $\pm$ 0.010 fm
\cite{Zhan:2011ji}, both from scattering measurements, confirmed the puzzle.
The situation has been discussed extensively in a number of papers---here we point out a review paper in
Annual Review of Nuclear and Particle Science \cite{Pohlreview:2012}---and in many talks and three dedicated workshops
\cite{prpw:2012,prpw:2014,prpw:2016}. 
It was generally agreed that new data were needed to resolve the puzzle, and a number of experiments have subsequently been developed. The MUon Scattering Experiment (MUSE) addresses the radius puzzle in a unique way, simultaneously measuring electron and muon scattering. A technical overview of the experiment is presented in~\cite{MUSE_TDR}.

The experiment in the PSI PiM1 channel \cite{PiM11,PiM12} operates with an approximately 3.5~MHz mixed secondary beam of electrons, muons, and pions.
This rate, combined with a planned 12 months of production data taking, provides sufficient luminosity and the needed statistics for the planned measurements.
The High Intensity Proton Accelerator provides a beam with sub-nanosecond bunch length and an RF frequency of 50.6 MHz. Combined with the nearly 23-m length of the PiM1 channel, this provides an adequate separation in particle arrival time relative to the accelerator RF phase---a few ns---for beam particle identification at three momenta, 115, 160, and 210 MeV/$c$, the momenta chosen for MUSE.
The nature of the PiM1 beam in the MUSE experiment---the RF timing for the mix of particles, the beam flux, and the few cm size of the beam spot---necessitates that beam particle times and trajectories be precisely measured.
Particle trajectories are determined with a GEM chamber telescope \citep{Milner_2014, Liyanage_2019}.

The technology adopted to address beam particle timing measurements uses plastic scintillators read out with Silicon photomultipliers (SiPMs) \citep{Simon_2019, Renker_2006, Brunner_2014}.
Three beam line detectors: the beam hodoscope (BH), beam monitor (BM), and beam focus monitor (BFM), were constructed using this technology.
The detectors were designed and fabricated at PSI.
This paper describes the major components of the detectors, tests and test results, and demonstrates the successful operation of these detectors.

\section{Beam Hodoscope}
\label{sec:BH}
The beam hodoscope is installed in the beam line at the most upstream end of the MUSE apparatus. 
The BH detects beam particles, to identify the particle type, determine the beam flux, and provide a starting time for time-of-flight (TOF) measurements.

Particle types are identified with RF time, the difference in time of the particle in the detector from the accelerator RF signal. 
This is essentially a TOF of the particle through the PiM1 channel, with an arbitrary offset, modulo the accelerator RF period.
The identification is performed with the full detector resolution in the event analysis, and, with reduced resolution, in a first-level particle identification (PID) trigger.

The beam flux is determined by two techniques.
First, scalers count the logic pulses, generated by discriminators, that go to both the trigger and multi-hit TDCs.
Counting these signals determines the total flux without distinguishing between particle types.
The first-level PID triggers are also scaled to count the flux of electrons and muons.
Second, although all of the signals from an event occur within a time window that is a few tens of nanoseconds wide, we record times of all signals in our TDCs over a much wider 1.5 $\mu$s TDC window, to study backgrounds including effects of randomly coincident beam particles. The BH TDCs then typically record 5 other beam particles in each event. The particle types of each can be identified using RF timing. With a 2 kHz trigger rate, the BH TDCs sample about 0.3\% of the beam (2 kHz $\times$ 1.5 $\mu$s time window), determining the particle distributions with statistical precision of order 1\% each second.

Two types of TOF measurements are performed.
In the event data analysis, the TOF of scattered particles from the BH to the scattered particle scintillators distinguishes between different reaction types. There are also dedicated measurements of TOF that are used to determine the distribution of momentum of muons and pions in the beam.

The electrons in the experiment are highly relativistic, with speed $\beta$ = $v/c$ $\approx$ 1. This allows the electron TOF to be used, in conjunction with tracking, to calibrate detector time offsets and measure the performance of the TOF system. For particles other than electrons, the incoming and scattered particle speeds are less than $c$ and the outgoing speed depends on scattering angle. Also, particle decays need to be taken into account. Thus, a more involved analysis is required to also use these data for calibrations.

The BH paddles also provide position information with several mm precision, which aids in identifying beam particle tracks.

{\em Requirements:}
Timing requirements arise from both the RF time and TOF measurements.
The RF time peaks are 300 - 400 ps rms, resulting from the distribution of protons on the M1 production target and the variation in flight paths through the PiM1 channel.
Requiring that the detector only minimally impacts the width of the RF peaks leads to a time resolution requirement better than approximately 150 ps.
TOF provides a more stringent resolution requirement of roughly 100 ps, for reaction identification at the highest beam momentum.
An efficiency of 99\% is required to efficiently collect data and reject backgrounds.
A rate capability above 3.5 MHz is needed to handle the beam flux.
The beam size at the BH leads to an active area requirement of 10 $\times$ 10 cm$^2$.
The detector must also determine position at the several mm level, for use in conjunction with GEM tracking chambers that immediately follow the BH in the beam line.
Finally, a thin detector, of order 0.5\% radiation length ($L_{rad}$), is needed to minimize effects on beam properties, including enlarging the beam spot on the target.

{\em Detector design:}
The detector design uses multiple planes of scintillator paddles,  
with SiPM read out, to satisfy the time resolution, rate capability, and position resolution requirements.
The planes alternate between horizontal and vertical orientations to better localize particle positions.
Two to four planes, each able to time particles at the approximately 100 ps (rms) level, are used 
depending on beam momentum. 
More planes are used at the higher momenta, as shown in Table~\ref{table:BHplanes}, where the TOF 
resolution requirements are more strict, but there is less multiple scattering of the beam per 
scintillator plane.
The upstream planes are slightly offset from the downstream planes so that the gaps between paddles 
do not line up.

\begin{table}[tbh]
\noindent\caption{Description of BH telescope. Position A is furthest upstream. Planes are positioned at 2-cm intervals along the beam line. The paddles are shifted perpendicular to the bar orientation by the offset given in the table.}
\vspace*{2mm}
\label{table:BHplanes}
\noindent{\footnotesize
\begin{tabular}{|l|l|r|l|}
\hline
Position & Orientation  & Offset & Beam momenta \\
      &             &  (mm)  & (MeV/$c$) \\
\hline
\hline
A & Horizontal & $2$ & 210\\
\hline
B & Vertical   & $-1$ &  210, 160\\
\hline
C & Horizontal & $0$ & 210, 160, 115\\
\hline
D & Vertical   & $1$ &210, 160, 115\\
\hline
\end{tabular}
}
\end{table}

We note that the initial BH design concept was based on scintillating fibers read out with multi-anode photomultipliers \cite{Erez:2016}. 
The design described here is a significant upgrade, yielding improved time resolution and
efficiency with a detector about half as thick. 
Each BH plane is approximately 0.5\% $L_{rad}$, which adds roughly 12 mr multiple scattering, 
widens the 1.5 cm radius beam spot on the target by several mm, and causes about 0.5 MeV energy loss, 
depending on particle species and beam momentum.

A BH hodoscope plane comprises sixteen BC-404 plastic scintillator paddles, each 100 mm long 
$\times$ 2 mm thick, read out at each end with Hamamatsu S13360-3075PE SiPMs. BC-404 was chosen due to its fast, 0.7-ns, rise time and large light output.
The scintillator material and SiPM choices are detailed in \ref{sec:SSchoice}.
Radiation damage to the SiPMs, discussed in \ref{subsec:Radiation}, was also a consideration in the 
detector geometry.
The six central paddles, in the more intense central core of the beam, are 4 mm wide.
They are flanked on each side by five 8 mm wide paddles. 
Thus each BH plane covers a 104 x 100 mm$^2$ area. 
Using narrower central paddles in the core of the beam better localizes a large fraction of the beam 
particles while also keeping the rate in every paddle significantly below 1 MHz and lowers the probability
that two beam particles pass through the same horizontal or vertical elements and cannot be resolved.

The Hamamatsu S13360-3075PE SiPMs have higher rate capability than required. 
The 4 mm wide paddles are read out by one SiPM at each end, while 8 mm wide paddles are read out by two SiPMs at each end, connected in series. 
The SiPMs themselves are soldered to custom made printed circuit boards, from which the signal is sent 
through LEMO connectors to amplifiers, described in subsection~\ref{subsec:Amplifiers}.

Frames to hold all scintillator paddles together were designed and produced at PSI. 
To avoid any reflections, they are made from matte black anodized aluminum. 
To make the detectors light-tight, with minimal material in the beam, the arrays of paddles are covered on back and front with 50 $\mu$m Tedlar foil.

\begin{figure}[tbh]
\centering
\includegraphics[width=0.49\textwidth]{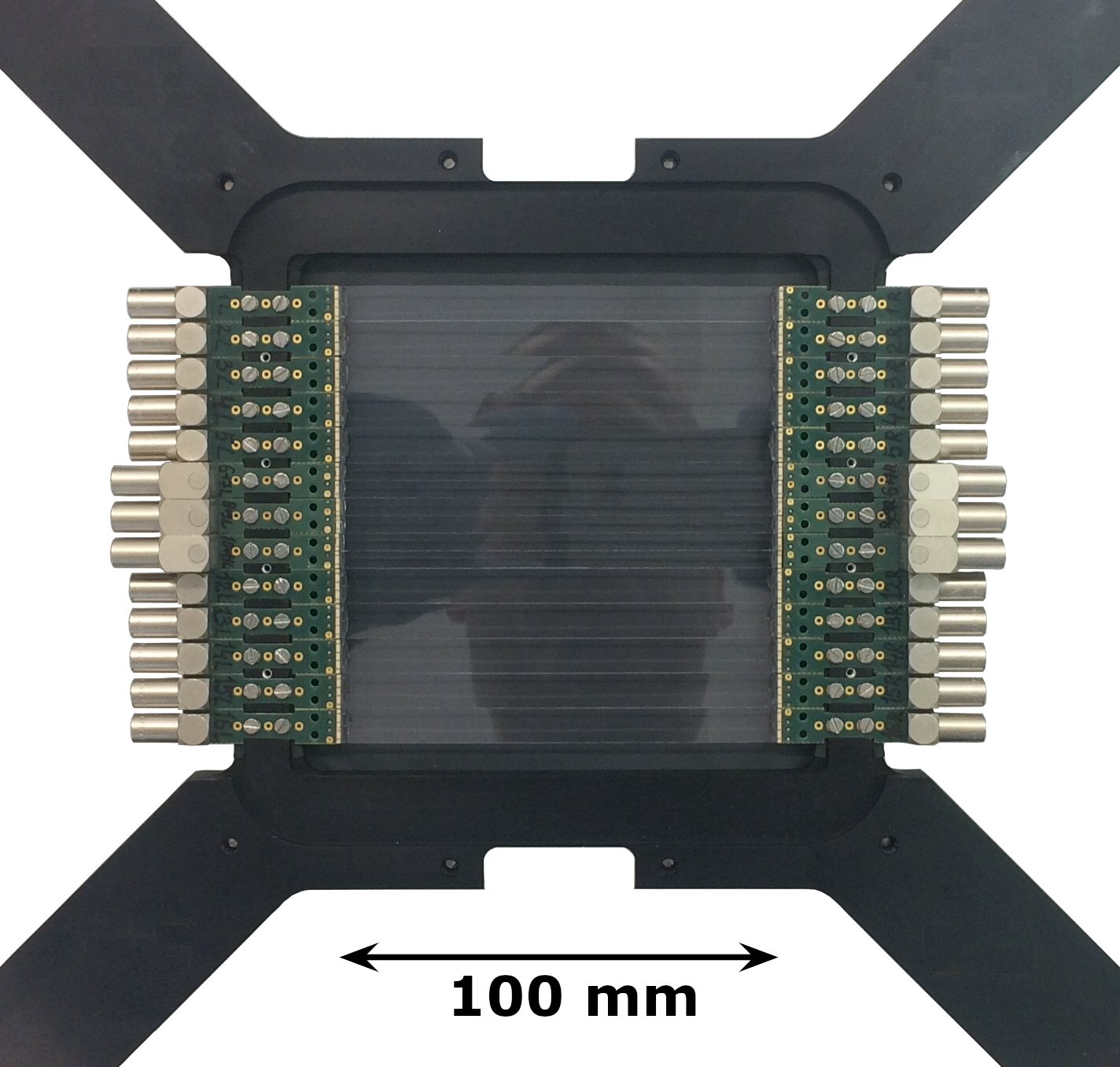}
\vskip-1mm\caption{A beam hodoscope plane during assembly. 
Visible are the transparent scintillator paddles, the black frame and the green SiPM carrier PCBs with LEMO connectors. 
While the outer paddles have individual PCBs, two neighbouring central paddles connect to one PCB.}
\label{BH_open}
\end{figure}
{\em Detectors built:}
Figure~\ref{BH_open} shows a photograph of a BH plane under construction.
Care must be taken in the construction process, in particular in the design of gluing fixtures and frames, so that the scintillator paddles are not mechanically stressed, which can lead to surface crazing and consequent reduced performance. 
This primarily impacted the construction of the 4-mm wide BH paddles.
A 6 $\mu$m air gap, established with an aluminized mylar foil used as a spacer during assembly, between the paddles efficiently suppresses optical cross-talk between them, but also causes a 0.15\% geometric inefficiency.\footnote{Analog signals from optical crosstalk are a few percent as large as the signals in the scintillator paddle struck. 
For sufficiently low thresholds this cross talk is easily detected. 
The discriminator threshold is chosen to maintain high (above 99\%) efficiencies with minimal (percent level) cross talk.}
During the prototyping, we tested coating the scintillator edges with Aluminum to suppress the cross-talk, but the coating damaged the scintillator surface, leading to a time resolution deteriorated by nearly 35 ps
(see Table~\ref{SiSi_Results} in \ref{sec:SSchoice}). 
 
\begin{figure}[tbh]
\centering
\includegraphics[width=0.49\textwidth]{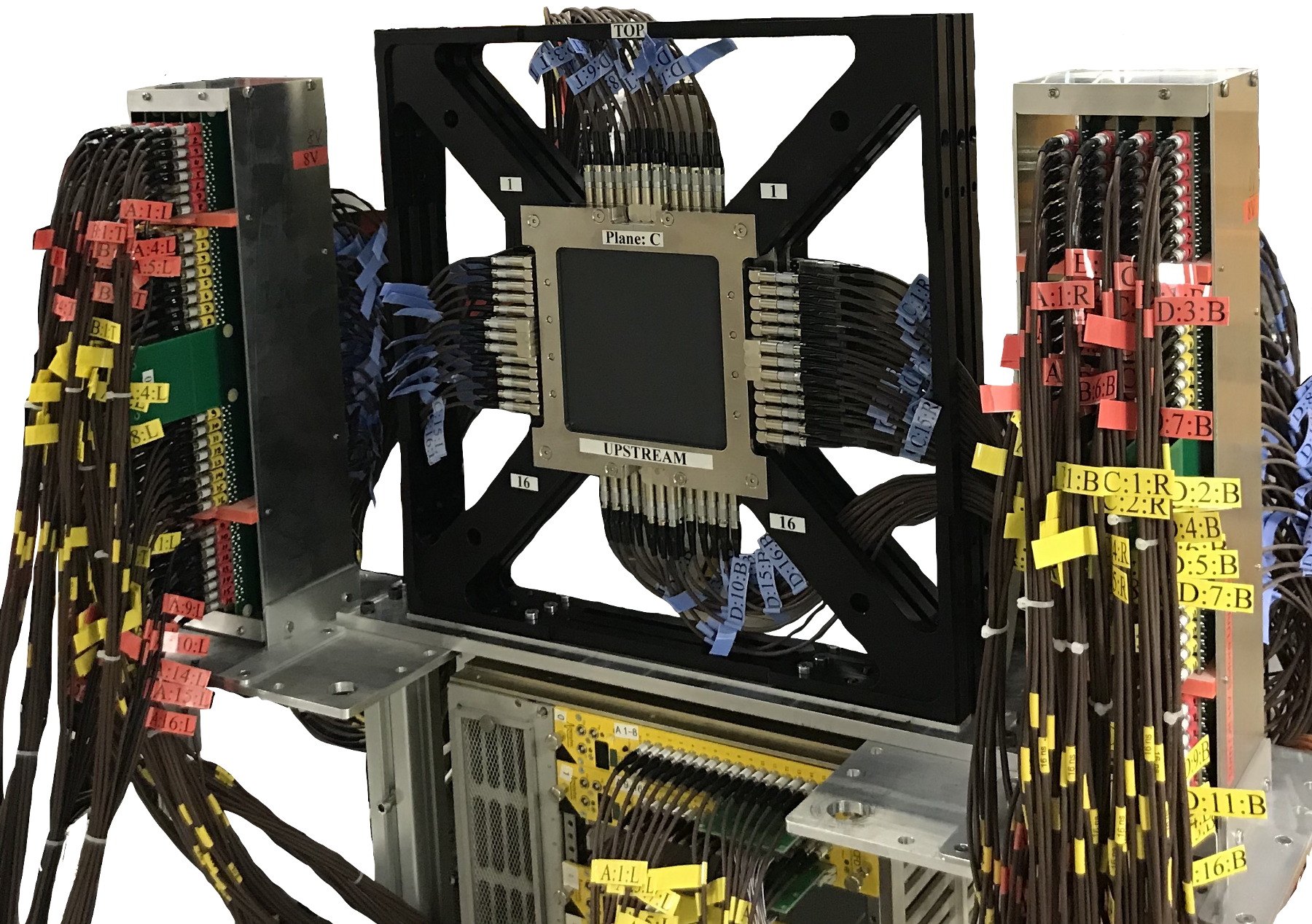}
\vskip-1mm\caption{Three Beam Hodoscope planes, together with Tel Aviv University (TAU) amplifiers to the sides and Mesytec CFDs below, installed in the MUSE apparatus.}
\label{3_BH_Planes}
\end{figure}

Four of the five BH planes were built as described above.
One plane was built with 13 8-mm wide paddles, while details of the assembly procedure were being adjusted to avoid surface crazing issues for the more delicate, 4-mm wide paddles.
All 5 planes were tested and exceeded experimental requirements; examples are shown in Section~\ref{sec:Results}. 
Figure~\ref{3_BH_Planes} shows a photograph of the MUSE setup with 3 BH planes installed, together with the amplifiers and read-out electronics, described respectively in subsections~\ref{subsec:Amplifiers} and \ref{subsec:Readout}.

\section{Beam Monitor}
\label{sec:BM}
The Beam Monitor (BM) is installed in the beam line downstream of the MUSE target system.
It provides a flux determination of beam particles downstream of the target.
It also provides a high-precision particle time measurement that is used in determining TOF from the BH to the BM. The dedicated TOF runs are used to measure the muon and pion momenta distributions.
Generally there should be no signal in the BM if there is a scattered particle. Signals in the BM indicate that there are forward backgrounds in the events from additional processes happening, beyond the elastic scattering event off the proton.
An example is M{\o}ller or Bhabha scattering, which can lead to a high-energy forward electron or positron
in conjunction with a low energy scattered particle, which might trigger the detector system.
Also, due to the roughly 1.5 $\mu$s long TDC window, the BM can be used in conjunction with the BH to detect randomly coincident, unscattered beam particles and identify their species.

{\em Requirements:}
The performance requirements for the scintillators and technology for the BM are the same as for the BH detector, but the geometry of the detector differs due to its position along the beam line, as described below.
Also, as with the BH, SiPMs and mechanical structures need to be away from the beam axis.
An added concern is that material in the beam causes back scattering of particles leading to background events in scattered particle detectors.

\begin{figure}[h]
\centering
\includegraphics[width=0.49\textwidth]{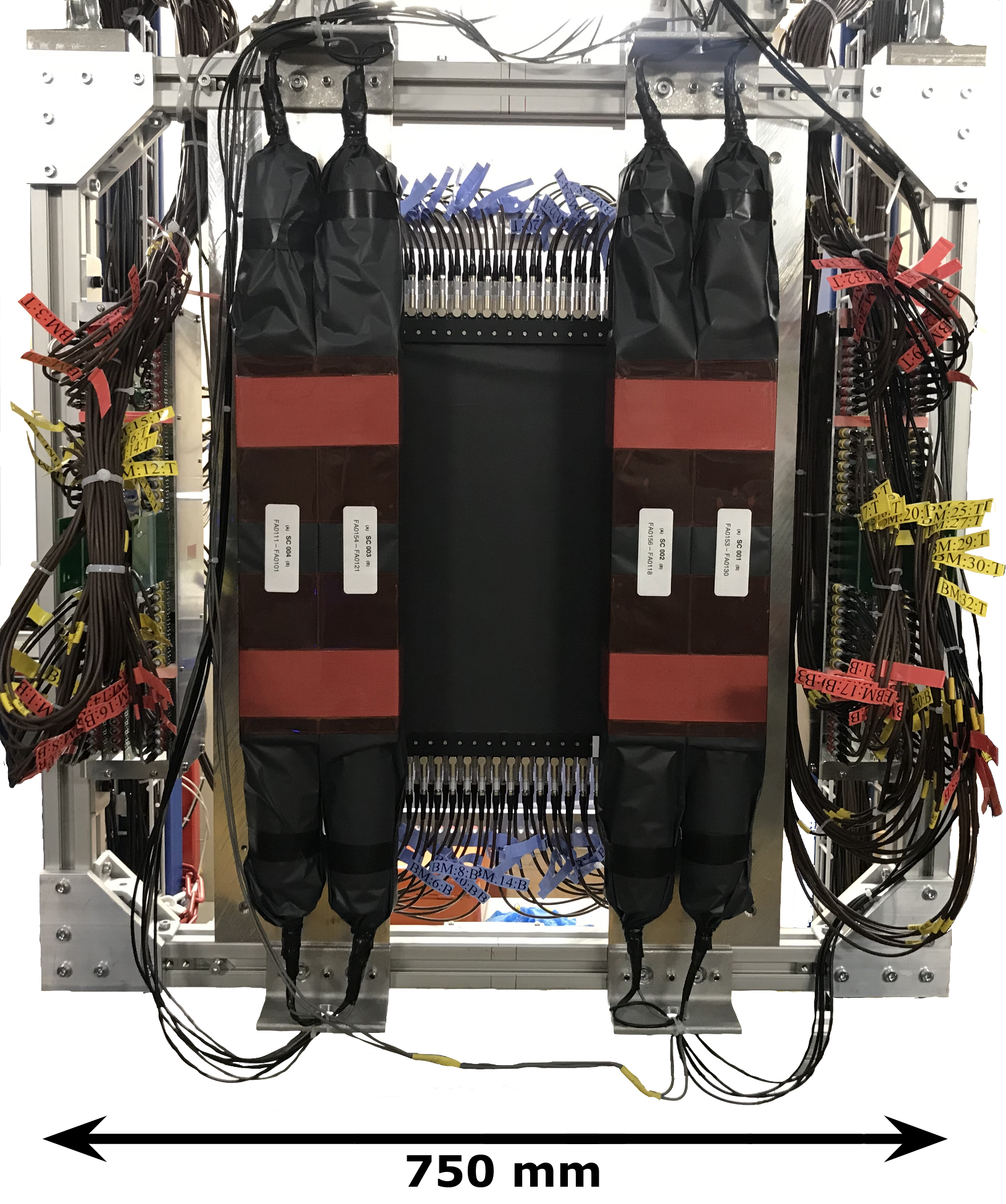}
\vskip-1mm\caption{Beam view of the Beam Monitor. The LEMO readout connectors of the two offset planes are seen, along with the bigger EJ-204 scintillators, and amplifiers to the sides.}
\label{fig:BM}
\end{figure}

{\em Detector design:}
Simulations show that background events can be efficiently suppressed if the BM covers an area with radius nearly 100 mm around the beam center.
It was decided to position the SiPMs 150 mm up and down from the beam center, to protect them from radiation damage.
The BM comprises a central scintillator hodoscope similar to that of the BH---see Section~\ref{sec:BH}---and, to enlarge the angular acceptance, four outer detectors.
The central hodoscope of the BM comprises two planes of 16-paddles, each 300 mm long $\times$ 12 mm wide $\times$ 3 mm thick BC-404 paddles.
To prevent alignment of the 6 $\mu$m gaps between two neighboring paddles, planes are 6 mm offset from each other.
The planes are 20.4 mm apart.
Each paddle is read out at each end by three Hamamatsu S13360-3075PE SiPMs in series.
To avoid optical cross-talk between two planes, there is a 50 $\mu$m Tedlar foil between them.
The outer four detectors are built of 30 cm long $\times$ 6 cm wide $\times$ 6 cm thick EJ-204\footnote{Eljen EJ-204 and Saint-Gobain BC-404 scintillators have similar properties. Both have a light output of 68\% Anthracene, 408 nm wavelength of maximum emission, 0.7 ns rise time, 1.8 ns decay time, 160 cm attenuation length, and 1.023 g/cm$^3$ density.} scintillators, read out at both ends with Hamamatsu R13435 PMTs.
The detectors of the central hodoscope is covered by 50 $\mu$m Tedlar on back and front to ensure that they are light-tight.
The frame is hung from a rail system, allowing movement to fixed, doweled positions, for differential TOF measurements.
A picture of the BM is shown in Figure~\ref{fig:BM}.

{\em Detectors built:}
The central hodoscope and large side paddles of the BM were fully assembled, installed, and commissioned. 
The detector has been used both as described above, and in some time of flight measurements with the large side paddles slid in from their normal flanking positions to be in front of the central hodoscope.
Achieved results are discussed in Section~\ref{sec:Results}.

\section{Beam Focus Monitor}
\label{sec:BFM}

The Beam Focus Monitor (BFM) is installed at the bottom of the liquid hydrogen target 
ladder inside the target vacuum chamber, described in \cite{LH2_target}. 
It is moved into the beam by the target slow controls in special runs to check the beam focus at the target position.
Comparing the rates in different channels of this detector gives us an indication of the quality of the beam spot horizontal focus at the target position.
The vertical beam profile can be mapped out more precisely by moving the MUSE target ladder in the vertical direction.
The BFM is also used in conjunction with the beam GEM telescope to check and calibrate the projection of the beam particle trajectories from the telescope to the target. 
The GEM telescope is used in normal running conditions to determine the beam spot at the target.

{\em Requirements:}
In order to directly sample the beam spot at the MUSE target, a compact detector with mm-sized scintillators, significantly smaller than the beam spot and similar in size to the resolution of the projected GEM tracks, is needed on the target ladder inside the target vacuum chamber~\cite{LH2_target}. 
Precise timing is not needed from the detector.

\begin{figure}[h]
\centering
\includegraphics[width=0.49\textwidth]{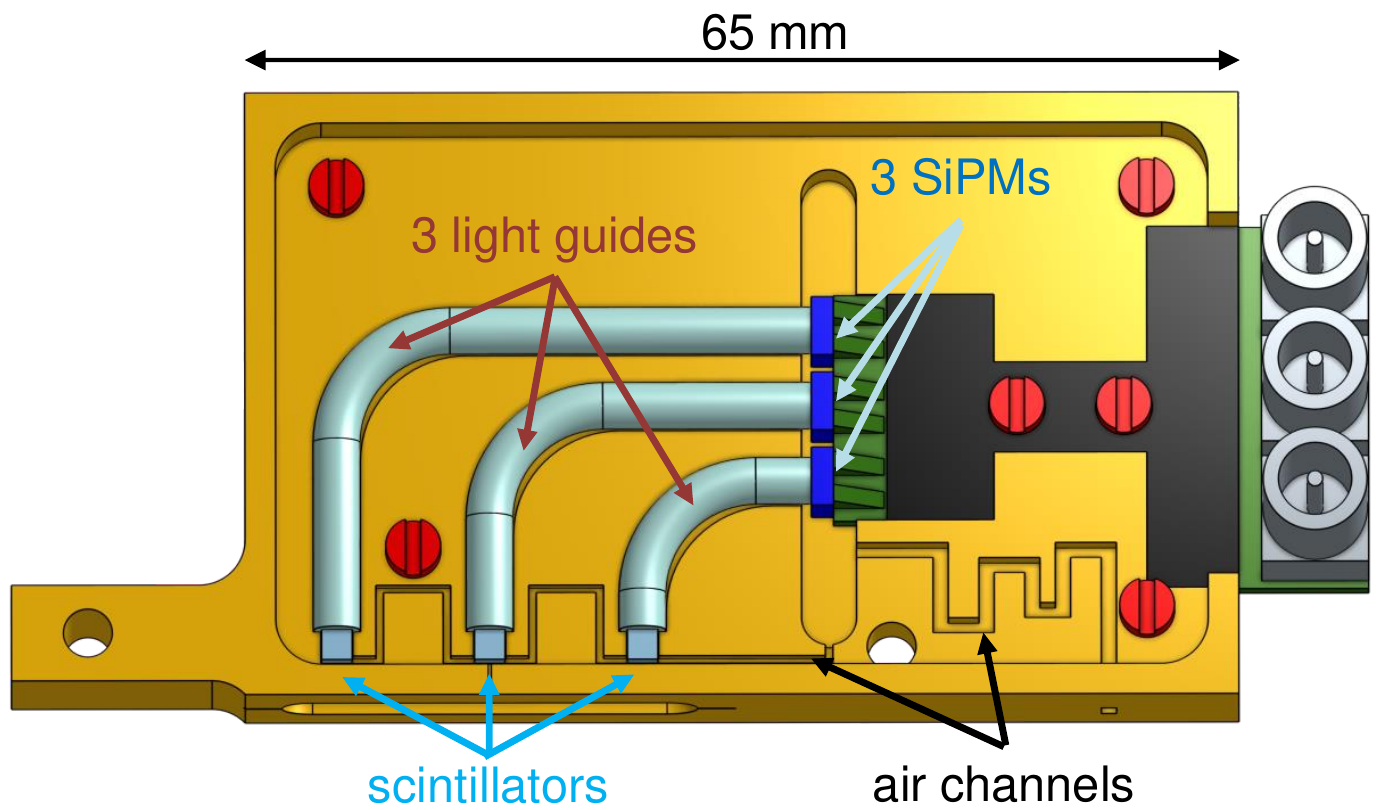}
\vskip-1mm\caption{A schematic view of the Beam Focus Monitor from Ref. \citep{LH2_target}. 
The beam profile is measured with three 8 mm$^3$ BC-404 scintillators. 
Light is transported with light guides to SiPMs, which are read out independently.}
\label{BFM_sketch}
\end{figure}

\begin{figure}[h]
\centering
\includegraphics[width=0.49\textwidth]{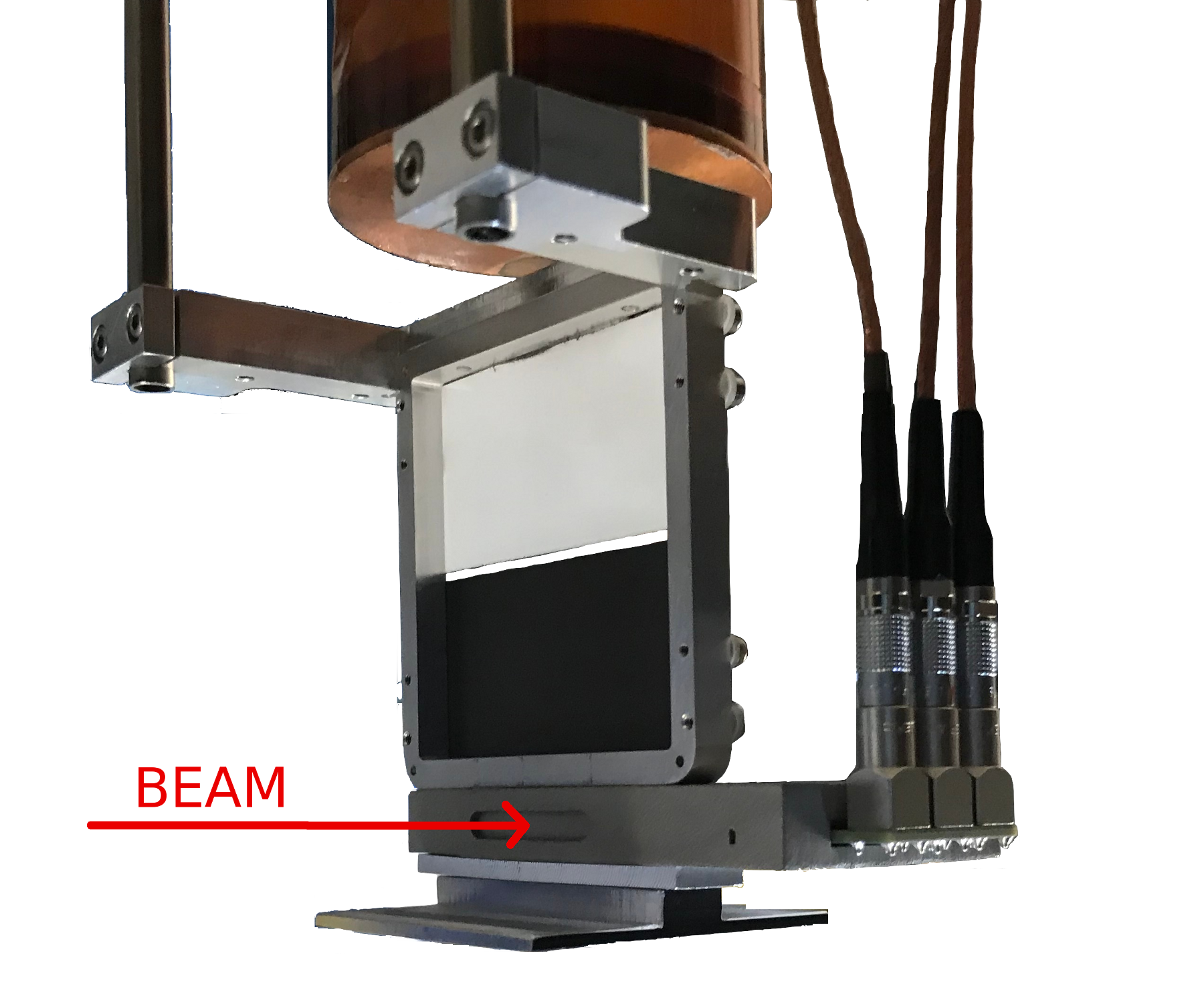}
\vskip-1mm\caption{Beam Focus Monitor installed at the lower end of the MUSE target ladder. Also visible are, from bottom to top, a plate with optical marks for alignment (below the BFM), carbon and plastic targets, as well as the kapton window of the hydrogen target cell.}
\label{BFM_ladder}
\end{figure}

{\em Detector design:}
A sketch of the BFM is shown in Figure~\ref{BFM_sketch};
Figure~\ref{BFM_ladder} shows the BFM mounted on the MUSE target ladder.
The Beam Focus Monitor (BFM) consists of three scintillator cubes mounted along a horizontal line, with 1 cm separation. 
The scintillators are made of 2 x 2 x 2 mm$^3$ Saint-Gobain BC-404 plastic.
They are connected by 3-mm diameter Saint-Gobain BCF-98 SC light-guides to Hamamatsu S13360-3050PE SiPMs.
To protect SiPMs from being directly in the beam line, the light guides are bent 90$^{\circ}$ to one side.
Each SiPM is read out independently.
The signals from the three SiPMs are taken out from the MUSE targets vacuum chamber via LEMO SWH.00.250.CTMPV feedthroughs.

The BFM frame was 3D-printed from an aluminum alloy (AlSi10Mg) using the Direct Metal Laser Sintering technique.
It is approximately 80 mm x 40 mm x 10 mm in outer dimensions.
The 3D-printed frame was successfully tested for light-tightness.
An air channel allows the device to be pumped out along with the vacuum chamber.
In testing a $10^{-7}$ mbar vacuum was reached in 28 hours.

{\em Detectors built:}
The BFM as described above was assembled and tested with beam. 
The signals from the three SiPMs showed the beam horizontal distribution---see Figure~\ref{BFM_beam_profile}.

\begin{figure}[h]
\centering
\includegraphics[width=0.49\textwidth]{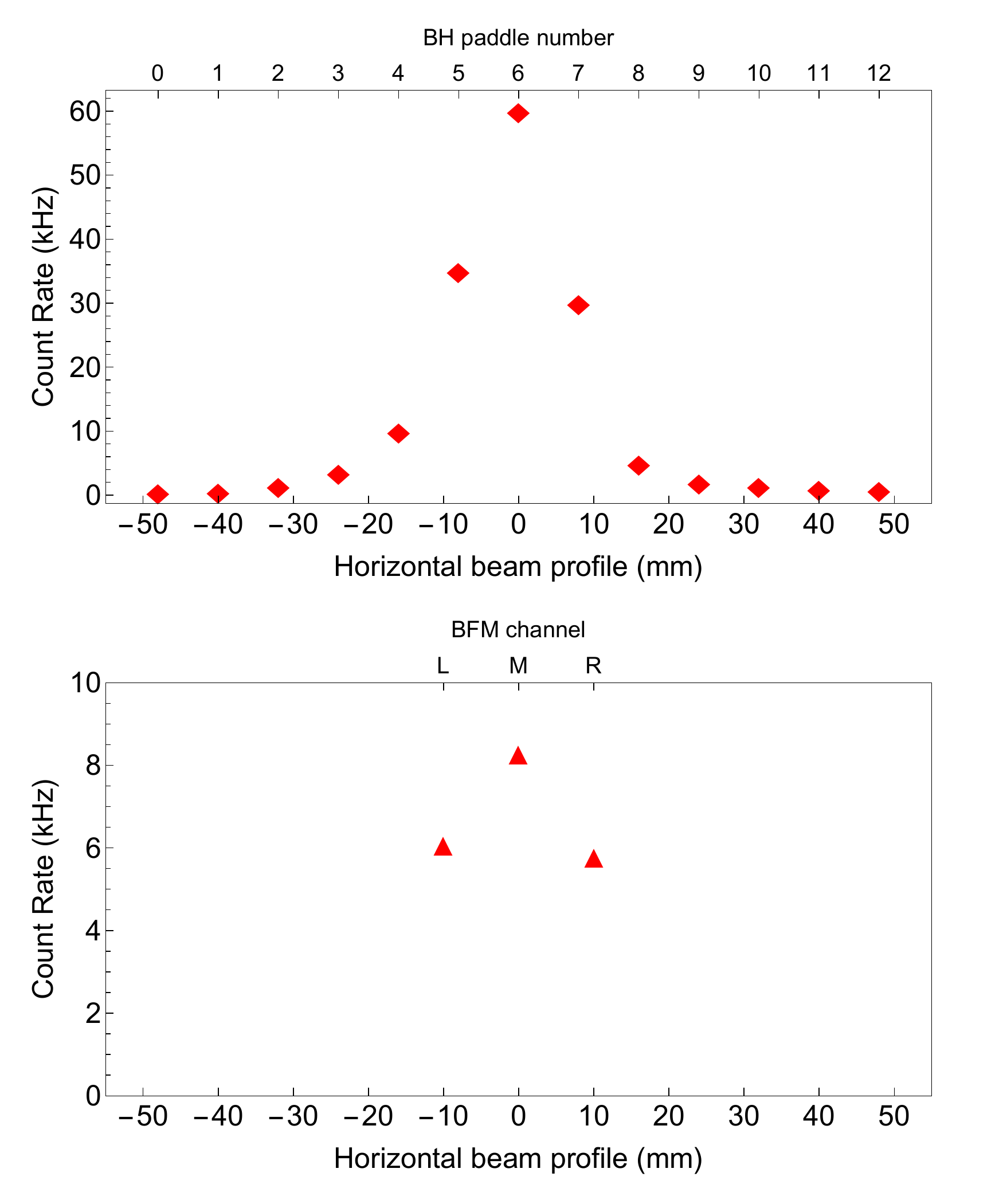}
\vskip-1mm\caption{Horizontal beam profiles from the BH and BFM detectors.
Top: BH profile of the beam about 34 cm upstream of the target center. 
The 13-paddle plane D was used.
Bottom: Three-channel BFM profile of the beam at the target position. 
}
\label{BFM_beam_profile}
\end{figure}

\section{Readout}
\label{sec:Readout}

\subsection{Amplifiers}
\label{subsec:Amplifiers}
\begin{figure}[h]
\centering
\includegraphics[width=0.49\textwidth]{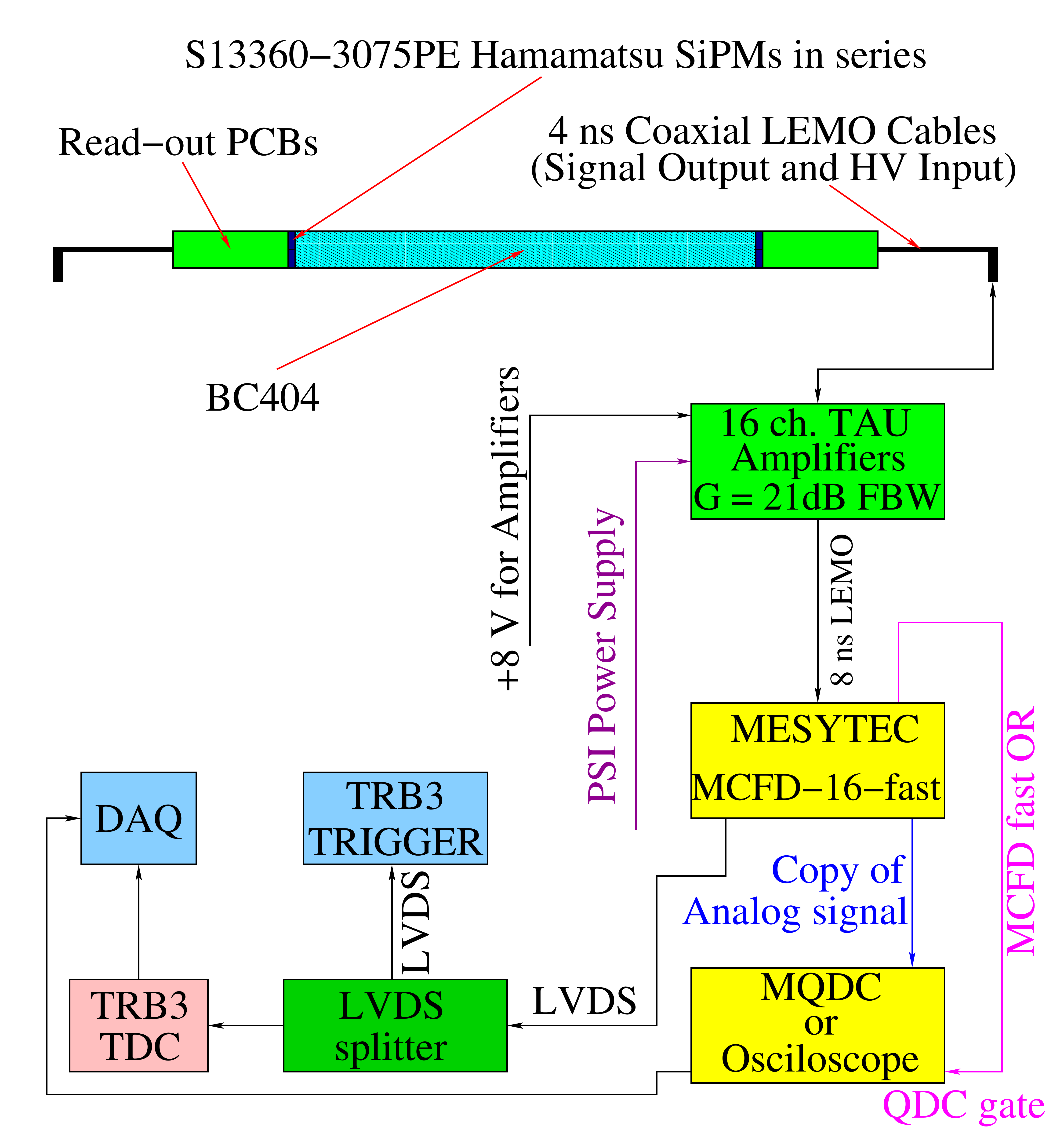}
\vskip-1mm\caption{Block diagram for a single channel of BH, BM and BFM. The SiPMs are read-out by a special-built amplifier constructed by the TAU group. A Mesytec MCFD discriminates the signal.
The CFD produces a time-delayed analog copy to be in sync with the CFD fast or output for subsequent digitization by the Mesytec QDC.
The CFD discriminated output is split with an LVDS splitter and fed into the TRB3 trigger system for experiment trigger and a TRB3-based TDC for timestamping.}
\label{fig:Readout}
\end{figure}

\begin{figure}[h]
\centering
\includegraphics[width=0.49\textwidth]{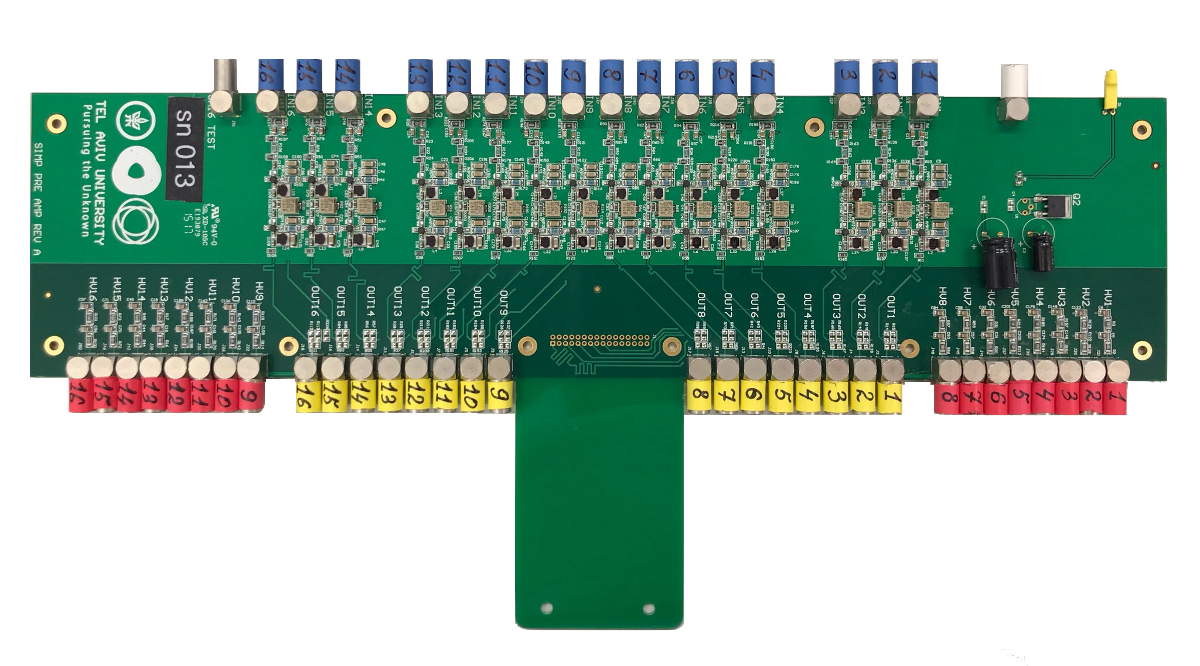}
\vskip-1mm\caption{The SiPM signal amplification circuit of Urs Greuter (PSI) implemented into a 16-channel card by Tel Aviv University.
The top-right connector is for the amplifier 8 V input. 
Top blue connectors are for the SiPM voltage input and SiPM signal read-out. 
The bottom connectors in red are the individual SiPM HV inputs, and in yellow are the amplified signal outputs.}
\label{16ch_Amplifier}
\end{figure}

\begin{figure}[h]
\centering
\includegraphics[width=0.49\textwidth]{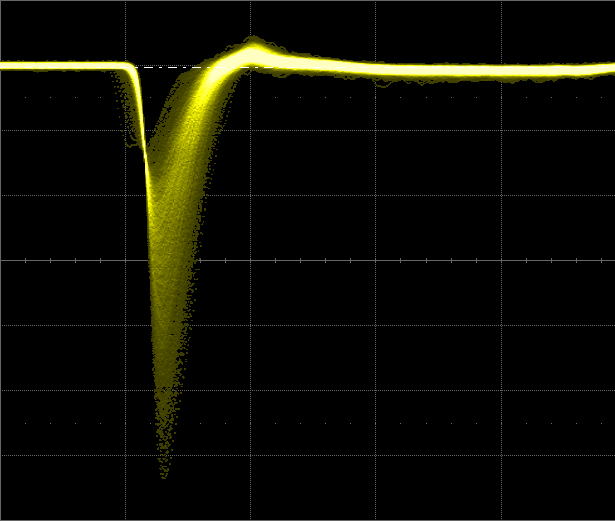}
\vskip-1mm\caption{Oscilloscope trace showing output signals from the SiPM amplifier resulting from illuminating 2-mm thick BC-404 with a $^{90}$Sr source. 
The scale is 200 mV/division vertical and 10 ns/division horizontal.
The output signal can be seen to return to the baseline within about 10 ns. The average rise (fall) time is 1.3 (3.3) ns.
}
\label{AmplifierOutput}
\end{figure}

A block diagram of the readout is shown in Figure~\ref{fig:Readout}. 
It consists of amplifiers and constant fraction discriminators, with analog signals sent to QDCs and discriminated signals sent to TDCs and trigger electronics.

Analog signals from SiPMs in all three detectors are amplified by the same type of amplifiers. 
These amplifiers follow the MAR-Amplifier design of Urs Greuter (PSI), shown in \ref{subsec:AmpCircuit}, as implemented on printed circuit boards designed and produced at Tel Aviv University (TAU). 
Figure~\ref{16ch_Amplifier} shows a photograph of a 16-channel amplifier card used for the BH and BM. 
A 3-channel version of the amplifier card was used for the BFM.
The amplified signal, shown in Figure ~\ref{AmplifierOutput},
has a 1.3 (3.3) ns rise (fall) time and typically a few hundred mV peak. 

\subsection{Readout Electronics}
\label{subsec:Readout}
Both the BH and BM are required to be high-precision timing detectors.
For this reason Mesytec Constant Fraction Discriminators (MCFD-16)~\cite{MCFD-16-fast} are used to discriminate the analog output from the amplifiers. 
These are 16-channel NIM-based modules.
The LVDS discriminator outputs from the MCFD-16s are sent to splitters, which directly couple the signals into multi-hit TRB3 TDCs~\cite{TRB3} for high-precision timing and copy them to a second TRB3 for triggering.

\begin{figure}[h]
\centering
\vspace*{-4mm}
\includegraphics[width=0.49\textwidth]{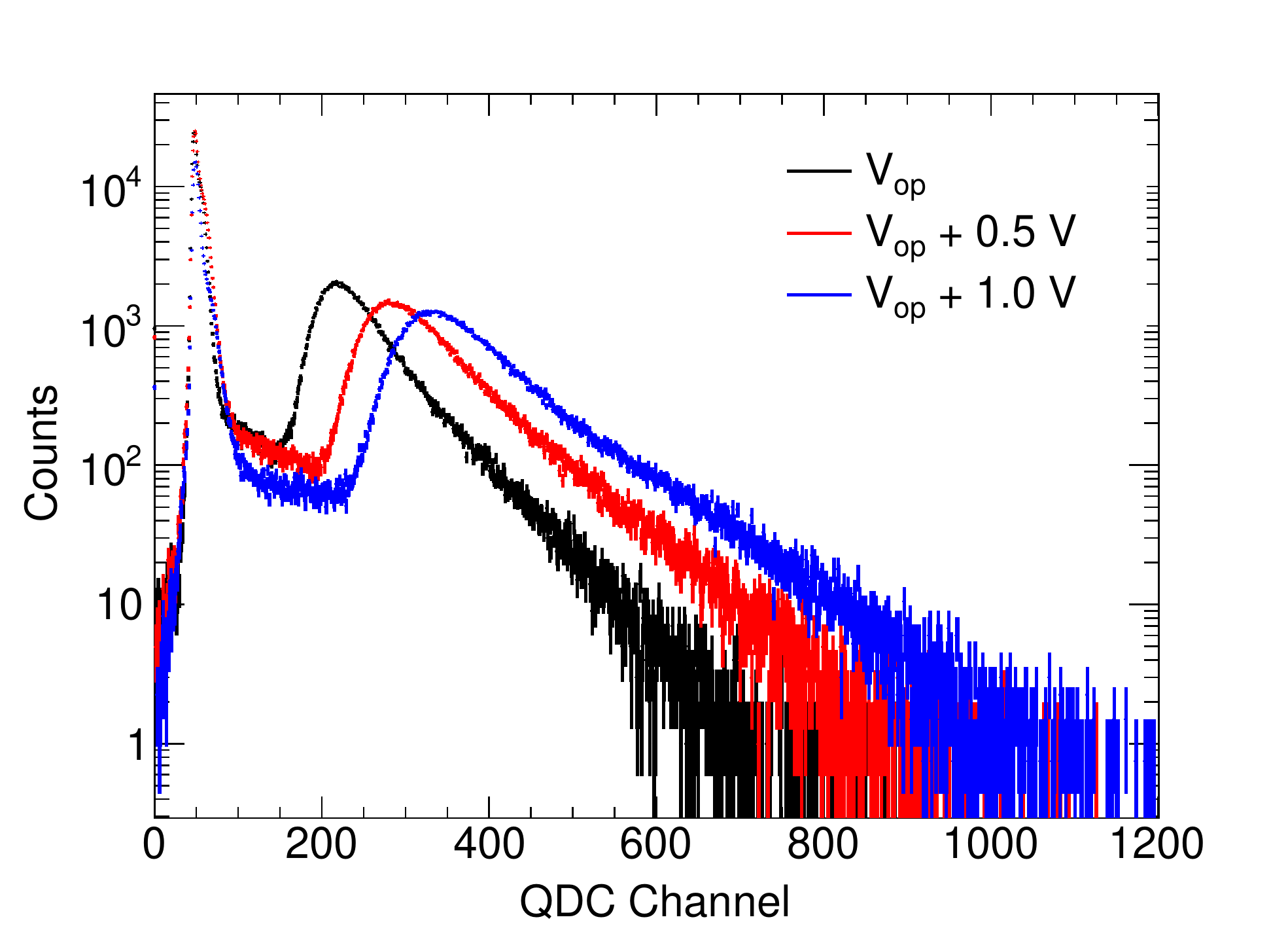}
\vskip-1mm\caption{  BH 8 mm wide paddle QDC spectra for different voltages applied to two Hamamatsu S13360-3075PE SiPMs in series. $V_{\text{op}}$ stands for factory recommended operating voltage, which is break-down voltage ($V_{\text{br}}$) + 3 V. The signal shape and behavior is the same for all BH paddles because $\approx 75\%$ of the area at the end of the paddles is covered by SiPMs, for both 8 mm wide paddles with two SiPMs and 4 mm wide paddles with one SiPM. }
\label{fig:BH_qdc_spectra}
\end{figure}

Detector HV, thresholds, and gains are monitored to keep detector preformance stable over time.
HV and thresholds are controlled through slow controls, while gains are monitored using QDC spectra.
The QDC spectra are generated using a second MCFD-16 output, a copy of the analog signal, that is sent to the 32-channel VME-based Mesytec Charge-to-Digital Converters (MQDC-32)~\cite{MQDC-32}.
The combined Mesytec MCFD plus MQDC system has a fast readout mode~\cite{Fast-Readout} that does not require additional delay of analog signals into the MQDC-32s.\footnote{A consequence of this choice is that, due to dead time in the QDCs, the charge is not read out for all events.}

The SiPM high voltages are tuned to allow a common threshold to be used.
Figure~\ref{fig:BH_qdc_spectra} shows that the BH QDC spectra are very sensitive to the SiPM input voltage.
The spectra were generated by irradiating a BH plane with a $^{90}$Sr source, with data acquisition triggered by a logical OR of the paddles in the plane.
The spectra shown include the peak from energy deposited by the source, a pedestal when other paddles triggered the data acquisition, and an intermediate region that has contributions from noise, crosstalk (light leakage from adjacent paddles), and reduced energy deposition when source particles clip a corner of the paddle or are randomly coincident but not fully in the QDC integration window. Because the BFM has only three channels, we opted for simplicity to use the same readout electronics as for the BH and BM.

\section{Performance Results}
\label{sec:Results}

\begin{figure}[h]
\centering
\includegraphics[width=0.49\textwidth]{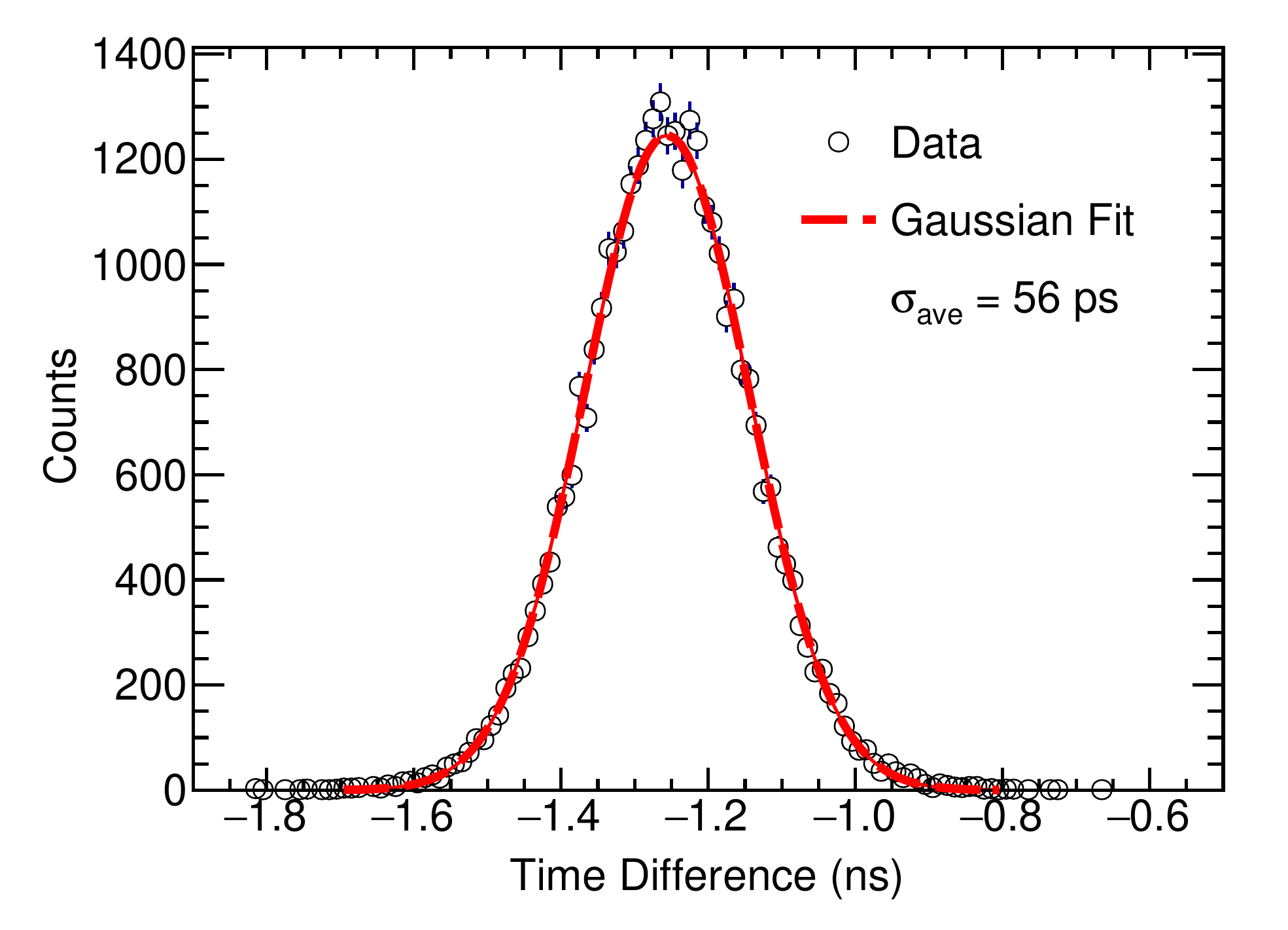}
\vskip-1mm\caption{Time difference between two ends of a paddle, for particles that passed through a 4-mm wide, centered, perpendicularly oriented paddle at a distance of 2 cm. 
Data were taken at 161 MeV/$c$ with positive polarity. 
Particles were 53\% $\pi$'s, 40\% $e$'s, and 7\% $\mu$'s.}
\label{fig:BH_time_resolution}
\end{figure}

BH and BM planes were tested with collimated $^{90}$Sr sources for quality control in construction and to measure the time resolution and also with beam.
Figure~\ref{fig:BH_time_resolution} shows an example of the time resolution for a BH paddle determined with beam.
Assuming that the intrinsic time resolution of the readout of both ends of a paddle are the same, then for a paddle illuminated by a point source the width of the time difference spectrum, $\sigma_{L-R}$, is roughly $\sqrt{2}$ times larger than the resolution of the readouts, and the width of the paddle mean time, $\sigma_{ave}$, is $\sqrt{2}$ times smaller than the resolution of the readouts.
As a result $\sigma_{\text{ave}}$ = $\sigma_{L-R}$/2.
Statistical uncertainties on the Gaussian fit are at the sub-picoseecond level.
We assign a $\pm$5 ps systematic uncertainty to BH and BM resolution measurements, from uncertainties due to the measurement technique and analysis assumptions, including non-Gaussian tails to the distributions we fit.

All five planes of BH were tested using the beam in PiM1 and the electronics described above. 
Typical time resolutions of $\sigma_t$ $<$ 100 ps, and a best of 55 ps, were achieved.
Efficiencies above 99.9\% were also achieved, with suitable thresholds, exceeding performance requirements (see Appendix \ref{subsec:Radiation}).
Figure~\ref{BHPA_Time_resolutions} shows as an example the test results of all 16 paddles for one of the BH planes.

It can be seen that the 8 mm wide paddles tend to have better resolution than the 4 mm wide paddles. This is largely due to the fact that the 8 mm wide paddles are easier to produce with high quality. For paddles of equal quality there should be no significant worsening of resolution of the wider paddles. For all paddles we cover $75\%$ of the area at the end of the paddle with SiPMs. The SiPMs on the wider paddles capture more direct light. For the reflected light the distance to a SiPM is the same independent of our paddle widths. The difference in distances to the different SiPMs on a paddle corresponds to 10 or 15 ps which does not significantly affect the resolution because it is small compared to the 700 ps rise time of the signal.

\begin{figure}[h]
\centering
\includegraphics[width=0.49\textwidth]{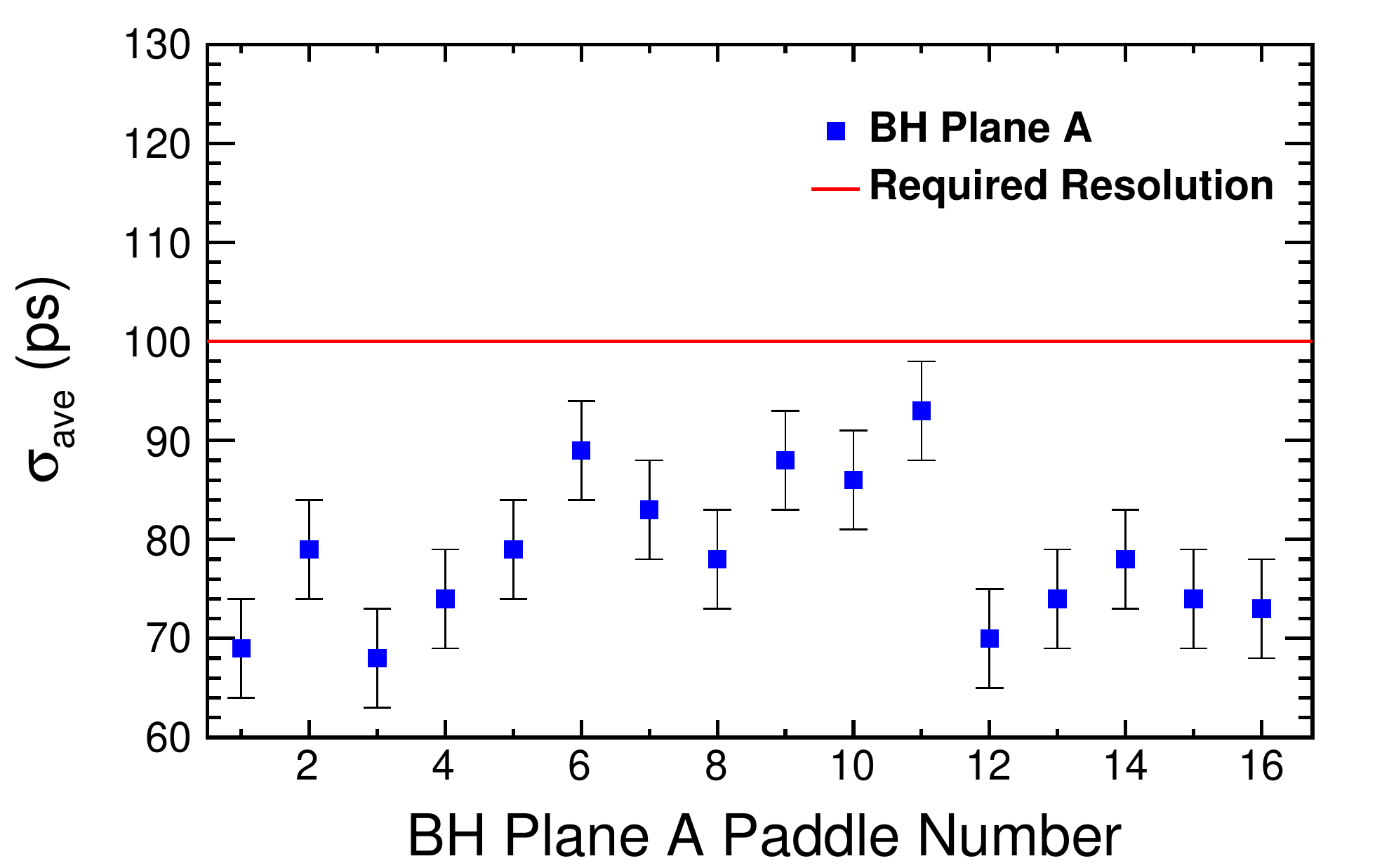}
\vskip-1mm\caption{Time resolution of all 16 BH-Plane-A paddles.}
\label{BHPA_Time_resolutions}
\end{figure}

{\em BM:}
The BM outer bars were built and tested at the University of South Carolina. 
A resolution of approximately 35 ps was obtained for each of them.
The BM central hodoscope paddles were tested with a centered $^{90}$Sr source, collimated by a 10-cm long, 5-mm diameter aluminum tube.
Data were obtained using an oscilloscope, which acted as a level discriminator with no walk correction. 
The position spread due to the collimator introduced an approximately 20 ps (rms) component to the time resolution.
A trigger in the oscilloscope was set to a coincidence between signals from both ends of the paddle, with both pulse heights above 50 mV. 
The results generally exceed the experimental requirements---see Figure~\ref{BM_Time_resolutions}.
Note that the resolution cannot be directly determined with beam data because there is no transverse detector near the BM that determines the position along the paddle; $\sigma_{L-R}$ reflects the beam size. Further study of the resolution (not shown) indicated that poorer resolutions, near 100 ps, are associated with paddles with smaller average amplified SiPM signal.

\begin{figure}[h]
\centering
\includegraphics[width=0.49\textwidth]{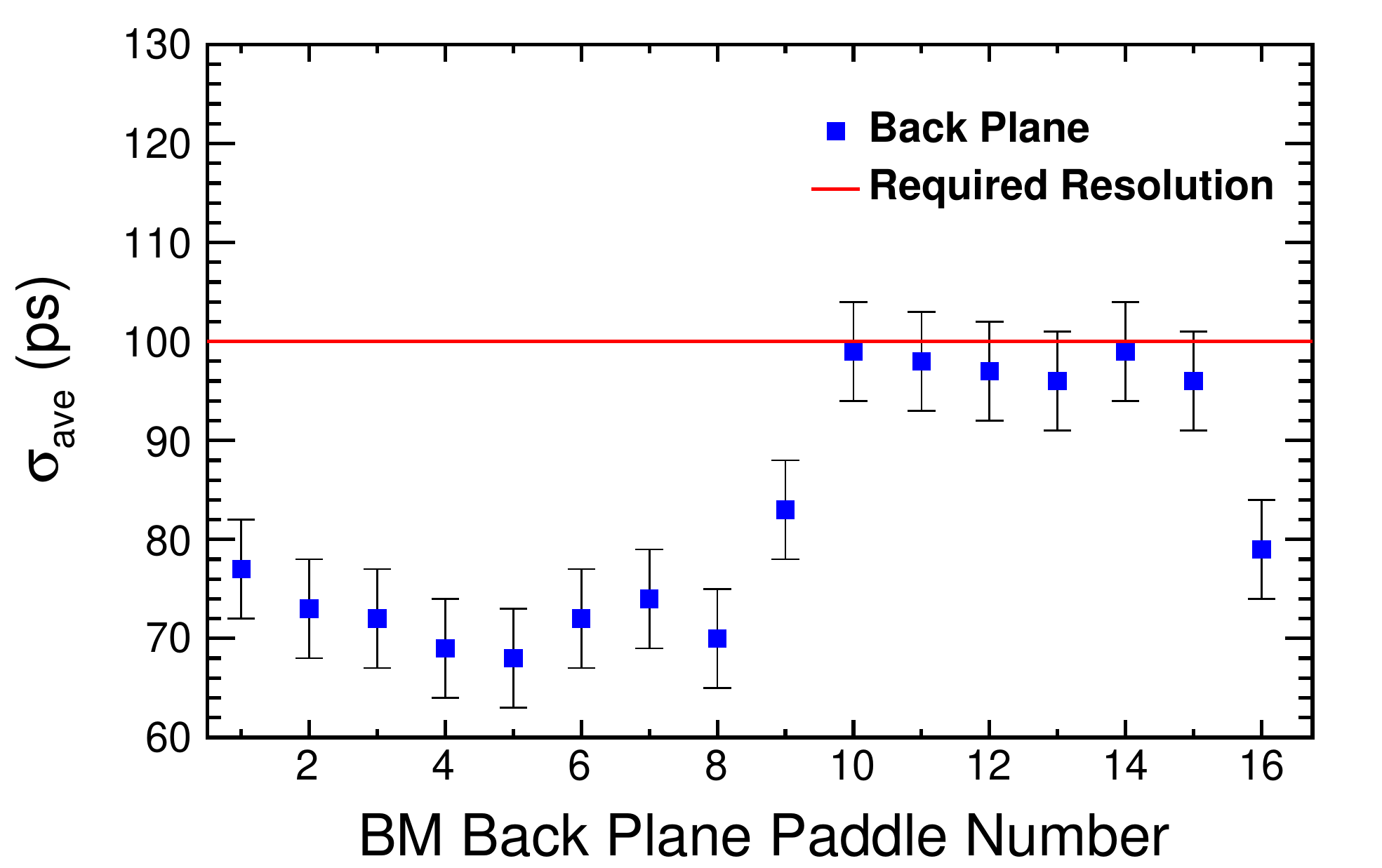}
\vskip-1mm\caption{Time resolution of BM back plane paddles. Paddles 1 - 8 used electronics channels on one 16-channel amplifier card, while paddles 9 - 16 used a second amplifier card.}
\label{BM_Time_resolutions}
\end{figure}

{\em BFM:}
The noise level for all three channels of BFM, after amplification, was found to be below 13 mV compared to a 400-500 mV signal.
Each channel was tested using a $^{90}$Sr source in front of each scintillator. 
The beam profile obtained with BFM is shown in the bottom part of the Figure~\ref{BFM_beam_profile}.

\begin{figure}[tbh]
\centering
\includegraphics[width=0.49\textwidth]{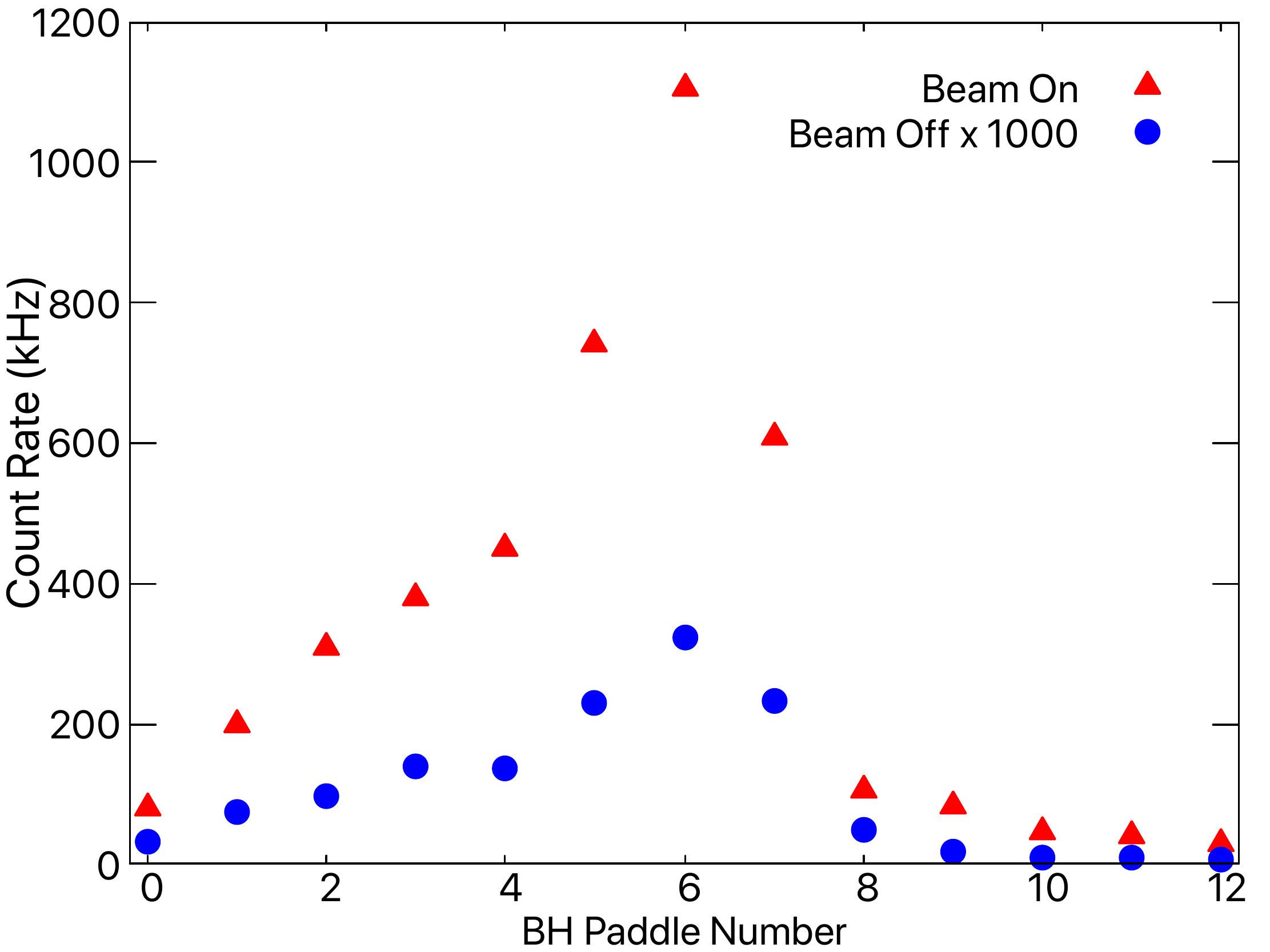}
\caption{Horizontal beam profile from instantaneous rates in BH plane D paddles just before and 3 minutes after beam is turned off. The beam off rate is multiplied by 1000 for ease of visualization.
}
\label{Beam_mimic}
\end{figure}
 
\begin{figure}[h]
\centering
\includegraphics[width=0.49\textwidth]{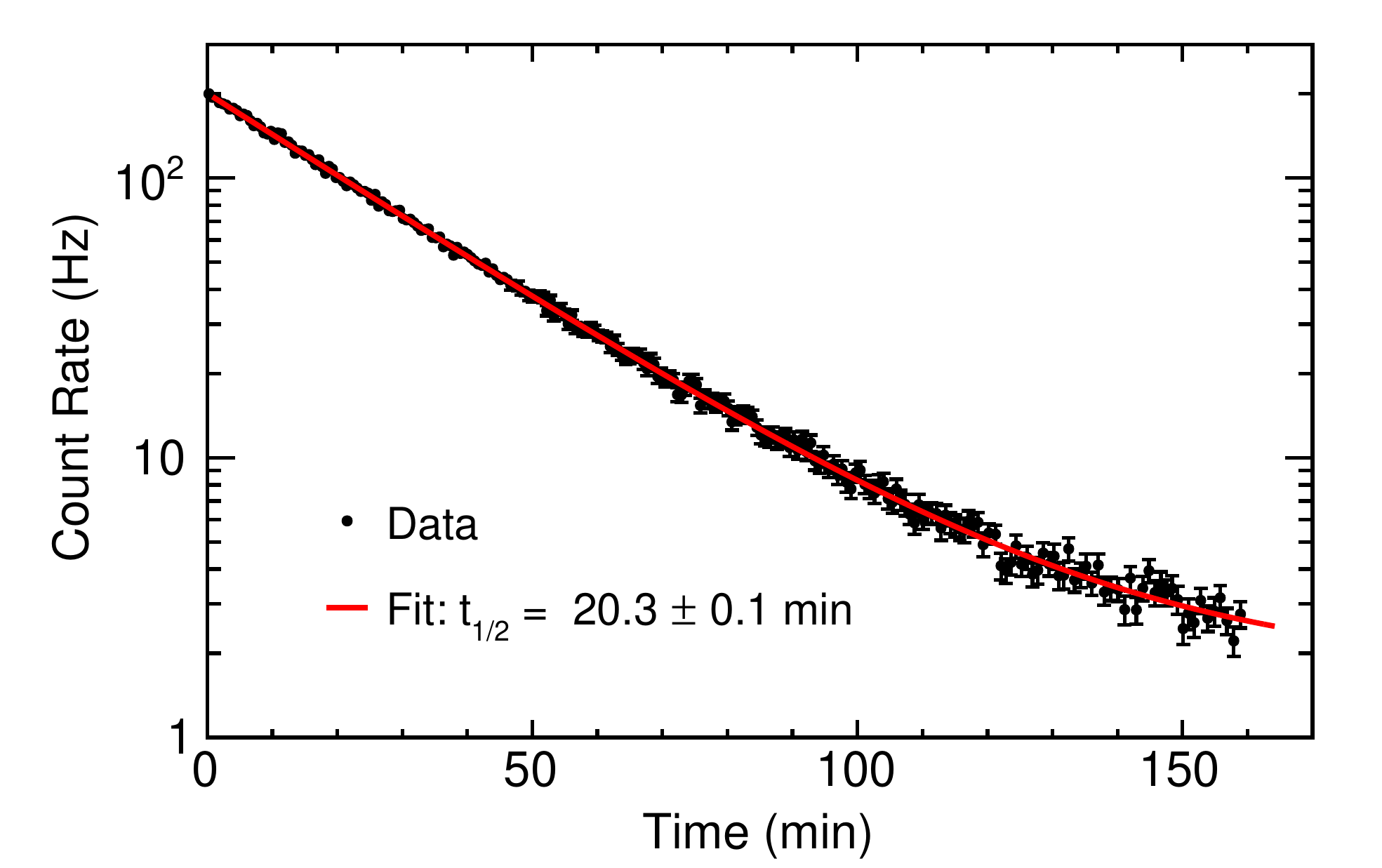}
\caption{Count rate from BH plane D paddle 7 vs. time after the beam was turned off. The data are fit with an exponential decay term plus a constant term to account for room background and should be compared to the $^{11}$C half life of $t_{1/2}$ = $20.364 (14)$ minutes.}
\label{Si_Activation}
\end{figure}
{\em BC-404 Activation:}
Activation of the scintillator material in the beam was observed. 
After the beam was turned off, the detector paddles still had an event distribution that mimicked the shape of the beam---see Figure~\ref{Beam_mimic}---and that gradually decreased with time.
The decrease of these rates was found to correspond to an exponential decay with a half-life 
consistent with that of $^{11}$C, which 
is $t_{1/2}$ = $20.364 (14)$ minutes,
plus a small constant room-background term.
It is likely that beam particles knock a neutron out of $^{12}$C nuclei in the plastic scintillator.

The intensity of the activation depends in detail on the beam history, since it requires stable running for several half lives to approach a steady state where production and decay rates are approximately equal. In intermittent running conditions, the activation is less. Also, the $^{11}$C production cross section would be expected to depend on beam momentum and incident particle type, which also varies with beam momentum. An upper limit for the activation of the BH during normal running can be extracted from the RF time spectrum, which shows a flat background underlying and between the $e$, $\mu$, and $\pi$ particle peaks at the level of a few tenths to one percent, depending on run conditions. This background includes a contribution from room background in addition to activation.

Plastic scintillators consist of about 91.5\% carbon and 8.5\% hydrogen, with density of 1.032 g/cm$^3$.
Taking into account the few cm$^2$ effective size of the beam at PiM1, we are irradiating approximately $10^{23}$ $^{12}$C atoms. 
Based on the decays observed during testing, at the roughly 3.5 MHz MUSE beam rate, we will produce
about 10$^8$ $^{11}$C per hour (28 kHz), or 10$^{12}$ $^{11}$C over the nearly 10$^4$ hours of the experiment. This is not a problem, either from the perspective of instantaneous background or material damage to the scintillator.

\section{Summary \& Conclusions}
\label{sec:Summary}
The Beam Hodoscope, Beam Monitor and Beam Focus Monitor detectors were built for the MUSE experiment at PSI, a measurement of the proton charge radius via elastic muon-proton and elastic electron-proton scattering. 
The experimental requires the BH and BM detectors to be more than 99\% efficient, and to have a time resolution of approximately 100 ps per plane, for beam rates up to 3.5 MHz.
This was accomplished with minimal effects on beam quality and background
generation by using 50-$\mu$m thick tedlar windows and 2-mm (3-mm) thick scintillator paddles for the BH (BM).
Both detectors used multiple planes of scintillator hodoscopes read out at each end with SiPMs.
An additional detector, the BFM, provides a measure of the beam at the target position where it comes to a focus, and allows calibration of the GEM telescope used in the scattering measurements.
All systems were designed with SiPMs positioned away from the beam to keep radiation damage at an acceptable level.
All three detectors were constructed and operated.
In all cases, the measurements demonstrated that the performance is sufficient for the MUSE physics goals.
The designs limit the radiation exposure so that the performance will be adequate over the length of the experiment.
Activation of the scintillator material was observed; this creates a manageable  background in the beam line detectors at the level of approximately 1\%.

\section{Acknowledgments}
\label{sec:Acknowledg}
We thank Alexey Stoykov (PSI) for the transfer of his experience in SiPM technology to us, and for his continuous help to the project.
We acknowledge PSI technicians Thomas Rauber, Manuel Schwarz and Florian Barchetti for their invaluable, continuous help and support.
We thank Urs Greuter for help with the readout amplifier design.
We acknowledge Paul Scherrer Institute for its hospitality and support.

This work was supported by the US National Science Foundation (NSF) grants 1614456, 1614773, 1614938, 1614850, 1714833, 1436680, 1505934, 1812402, 1807338, 1649909, and 1913653, 
by the United States-Israel Binational Science Foundation (BSF) grant 2012032 and 2017673, 
by the NSF-BSF grant 2017630, 
by the U.S. Department of Energy (DOE) with contract no.~DE-AC02-06CH11357, DE-SC0012589, DE-SC0016577, DE-SC0012485, DE-SC0016583, and DE-FG02-94ER40818,
by PSI, by Schweizerischer Nationalfonds (SNF) 200020-156983, 132799, 121781, 117601, by the Azrielei Foundation, by the Swiss State Secretariat for Education, Research and Innovation (SERI) grant FCS 2015.0594, and by Sigma Xi grants G2017100190747806 and G2019100190747806.

%\newpage
\appendix

%\section{SiPM and Scintillator choice}
\section{Scintillator and SiPM Selection}
\label{sec:SSchoice}
\renewcommand\thefigure{A.\arabic{figure}}    
\renewcommand\thetable{A.\arabic{table}}  
\setcounter{figure}{0}    
\setcounter{table}{0}

\begin{table*}[t]
    \caption{Time resolutions (ps) from tests with different combinations of SiPMs and scintillators.
EJ204c is EJ204 scintillator with aluminized coating.
BC422a used a 12 $\mu$m air gap, instead of 6 $\mu$m.
BC422p used a 12 $\mu$m gap between paddles with 6 $\mu$m air and 6 $\mu$m AlBoPET. Systematic error is of the order of 5 ps, the statistical error is negligible.}
\label{SiSi_Results}
\setlength\tabcolsep{1pt}
\footnotesize
\begin{tabular}{ccccccccccccccc}
\hline
\multicolumn{1}{|c|}{SiPM}                                                          & \multicolumn{4}{c||}{4mm wide, 100mm long}                                                                         & \multicolumn{6}{c||}{5 mm wide, 100 mm long}                                                                                                                                    & \multicolumn{4}{c|}{8 mm wide, 100 mm long}                                                                       \\ 
\multicolumn{1}{|c|}{}                                                          & \multicolumn{1}{c}{BC404} & \multicolumn{1}{c}{BC418} & \multicolumn{1}{c}{BC420} & \multicolumn{1}{c||}{BC422} & \multicolumn{1}{c}{EJ204} & \multicolumn{1}{c}{EJ204c} & \multicolumn{1}{c}{BC420} & \multicolumn{1}{c}{BC422} & \multicolumn{1}{c}{BC422a} & \multicolumn{1}{c||}{BC422p} & \multicolumn{1}{c}{BC404} & \multicolumn{1}{c}{BC418} & \multicolumn{1}{c}{BC420} & \multicolumn{1}{c|}{BC422} \\ \hline\hline
\multicolumn{1}{|c|}{\begin{tabular}[c]{@{}c@{}}S13360-\\ 3050PE\end{tabular}} & \multicolumn{1}{c|}{65}    & \multicolumn{1}{c|}{}      & \multicolumn{1}{c|}{}      & \multicolumn{1}{c||}{68}    & \multicolumn{1}{c|}{80}    & \multicolumn{1}{c|}{}       & \multicolumn{1}{c|}{77}    & \multicolumn{1}{c|}{60}    & \multicolumn{1}{c|}{77}     & \multicolumn{1}{c||}{75}     & \multicolumn{1}{c|}{67}    & \multicolumn{1}{c|}{}      & \multicolumn{1}{c|}{}      & \multicolumn{1}{c|}{53}    \\ \hline
\multicolumn{1}{|c|}{\begin{tabular}[c]{@{}c@{}}S13360-\\ 3075PE\end{tabular}} & \multicolumn{1}{c|}{61}    & \multicolumn{1}{c|}{}      & \multicolumn{1}{c|}{}      & \multicolumn{1}{c||}{63}    & \multicolumn{1}{c|}{}      & \multicolumn{1}{c|}{}       & \multicolumn{1}{c|}{}      & \multicolumn{1}{c|}{94}    & \multicolumn{1}{c|}{}       & \multicolumn{1}{c||}{}       & \multicolumn{1}{c|}{}      & \multicolumn{1}{c|}{}      & \multicolumn{1}{c|}{}      & \multicolumn{1}{c|}{78}    \\ \hline
\multicolumn{1}{|c|}{\begin{tabular}[c]{@{}c@{}}S12572-\\ 025P\end{tabular}}   & \multicolumn{1}{c|}{}      & \multicolumn{1}{c|}{}      & \multicolumn{1}{c|}{}      & \multicolumn{1}{c||}{}      & \multicolumn{1}{c|}{80}    & \multicolumn{1}{c|}{114}    & \multicolumn{1}{c|}{}      & \multicolumn{1}{c|}{99}    & \multicolumn{1}{c|}{}       & \multicolumn{1}{c||}{}       & \multicolumn{1}{c|}{}      & \multicolumn{1}{c|}{}      & \multicolumn{1}{c|}{}      & \multicolumn{1}{c|}{74}    \\ \hline
\multicolumn{1}{|c|}{AdvanSiD\begin{tabular}[c]{@{}c@{}}\ \\ \  \end{tabular}}                                                 & \multicolumn{1}{c|}{67}    & \multicolumn{1}{c|}{}      & \multicolumn{1}{c|}{}      & \multicolumn{1}{c||}{65}    & \multicolumn{1}{c|}{70}    & \multicolumn{1}{c|}{}       & \multicolumn{1}{c|}{79}    & \multicolumn{1}{c|}{88}    & \multicolumn{1}{c|}{}       & \multicolumn{1}{c||}{}       & \multicolumn{1}{c|}{}      & \multicolumn{1}{c|}{}      & \multicolumn{1}{c|}{}      & \multicolumn{1}{c|}{}      \\ \hline
                                                                               &                            &                            &                            &                            &                            &                             &                            &                            &                             &                             &                            &                            &                            &                            \\ \hline 
\multicolumn{1}{|c|}{SiPM}                                                          & \multicolumn{4}{c|}{4 mm wide, 161.5 mm long}                                                                     & \multicolumn{6}{c||}{5 mm wide, 161.5 mm long}                                                                                                                                  & \multicolumn{4}{c|}{8 mm wide, 161.5 mm long}                                                                     \\ 
\multicolumn{1}{|c|}{} & \multicolumn{1}{c}{BC404} & \multicolumn{1}{c}{BC418} & \multicolumn{1}{c}{BC420} & \multicolumn{1}{c||}{BC422} & \multicolumn{1}{c}{BC404} & \multicolumn{1}{c}{BC418}  & \multicolumn{1}{c}{BC420} & \multicolumn{1}{c}{BC422} & \multicolumn{1}{c}{}       & \multicolumn{1}{c||}{}       & \multicolumn{1}{c}{BC404} & \multicolumn{1}{c}{BC418} & \multicolumn{1}{c}{BC420} & \multicolumn{1}{c|}{BC422} \\ \hline\hline
\multicolumn{1}{|c|}{\begin{tabular}[c]{@{}c@{}}S13360-\\ 3050PE\end{tabular}} & \multicolumn{1}{c|}{65}    & \multicolumn{1}{c|}{}      & \multicolumn{1}{c|}{}      & \multicolumn{1}{c||}{}      & \multicolumn{1}{c|}{87}    & \multicolumn{1}{c|}{85}     & \multicolumn{1}{c|}{86}    & \multicolumn{1}{c|}{91}    & \multicolumn{1}{c|}{}       & \multicolumn{1}{c||}{}       & \multicolumn{1}{c|}{67}    & \multicolumn{1}{c|}{64}    & \multicolumn{1}{c|}{77}    & \multicolumn{1}{c|}{75}    \\ \hline
\multicolumn{1}{|c|}{\begin{tabular}[c]{@{}c@{}}S13360-\\ 3075PE\end{tabular}} & \multicolumn{1}{c|}{59}    & \multicolumn{1}{c|}{}      & \multicolumn{1}{c|}{}      & \multicolumn{1}{c||}{}      & \multicolumn{1}{c|}{}      & \multicolumn{1}{c|}{}       & \multicolumn{1}{c|}{}      & \multicolumn{1}{c|}{}      & \multicolumn{1}{c|}{}       & \multicolumn{1}{c||}{}       & \multicolumn{1}{c|}{}      & \multicolumn{1}{c|}{}      & \multicolumn{1}{c|}{}      & \multicolumn{1}{c|}{}      \\ \hline
\multicolumn{1}{|c|}{AdvanSiD\begin{tabular}[c]{@{}c@{}}\ \\ \  \end{tabular}}                                                 & \multicolumn{1}{c|}{72}    & \multicolumn{1}{c|}{}      & \multicolumn{1}{c|}{}      & \multicolumn{1}{c||}{}      & \multicolumn{1}{c|}{78}    & \multicolumn{1}{c|}{}       & \multicolumn{1}{c|}{}      & \multicolumn{1}{c|}{}      & \multicolumn{1}{c|}{}       & \multicolumn{1}{c||}{}       & \multicolumn{1}{c|}{65}    & \multicolumn{1}{c|}{}      & \multicolumn{1}{c|}{}      & \multicolumn{1}{c|}{}      \\ \hline
\end{tabular}
\end{table*}

Different scintillators have different light emission ranges and peak emissions. 
For example, Saint-Gobain BC-404 plastic scintillator emits light in the roughly 380 - 500 nm range, with peak emission at 408 nm.
For BC-422, the range is about 350 - 460 nm, peaking at 378 nm.

Different scintillators also differ in pulse rise and decay times, light attenuation length, etc. 
BC-404 has a pulse rise time of 0.7 ns and 2.2 ns
decay time, while BC-422 has a pulse rise time of 0.35 ns and 1.3 ns decay time.
BC-404 has a light attenuation length of 140 cm, while the BC-422 light attenuation length is very short.

SiPMs have different light detection efficiencies as a function of wavelength.
The peak photon detection efficiencies for AdvanSiD ASD-NUV3S-P-40, Hamamatsu S13360-3050PE and
Hamamatsu S13360-3075PE SiPMs are about 420 nm (43\%), 450 nm (40\%) and 450 nm (50\%), respectively, which roughly matches the wavelengths produced in the plastic scintillators.

Based on these observations, several prototypes, with various SiPM $+$ Scintillator combinations, were made and tested to optimize the time resolution and efficiency.
Eljen Technology EJ-204 and Saint-Gobain BC-404, BC-418, BC-420, BC-422 scintillators with different geometry
were tested in combination with four different SiPMs: AdvanSiD ASD-NUV3S-P-40, Hamamatsu S13360-3025PE,
Hamamatsu S13360-3050PE and Hamamatsu S13360-3075PE.
All tested scintillators were 2 mm thick, but 4 mm, 5 mm or 8 mm wide. 
The 4 and 5 mm wide prototypes had 1 SiPM at each end, while 8 mm wide paddles had 2 SiPMs at each end connected in series.
The BH planes need a 100 mm x 100 mm active area to extend out into the far tails of the beam.
Due to concerns about radiation damage to SiPMs, prototype paddles were made in 2 different lengths: 100 mm and, to have less ($\approx$ 4.4 times) radiation on SiPMs, 161.5 mm long. 
The results of these tests are shown in the Table~\ref{SiSi_Results}.

Results with AdvanSiD ASD-NUV3S-P-40, Hamamatsu S13360-3050PE and Hamamatsu S13360-3075PE SiPMs in combination with all tested scintillators agree within experimental uncertainties---estimated to be $\pm$5 ps, dominated by systematics---with each other, and exceed experimental requirements.
Results with BC-404 and BC-422 were slightly better than others. 
The BM requires 300 mm long paddles, so, because of BC-422's short attenuation length, a decision to use BC-404 in both detectors was made.

For the BM, three BC-404 prototype paddles with 300-mm length, 3-mm thickness, and 12-mm width were constructed.
The readout used 3 SiPMs attached in series at each end.
The efficiency was determined by sandwiching the prototypes between 2 smaller trigger scintillators, and seeing what fraction of events included a discriminated signal from the prototype.
Table~\ref{table:BM_prototypes} shows test results; see also Section~\ref{sec:Results}.

\begin{table}[h]
\noindent\caption{Time resolutions and efficiencies for 3 mm thick, 300 mm long and 12 mm wide BC-404 BM paddles. 
All results are better than the experimental requirements.}
\vspace*{2mm}
\label{table:BM_prototypes}
\noindent{\footnotesize
\begin{tabular}{|c|c|c|c|}
\hline
Scintillator & SiPM & {\large $\sigma$}{\small $_{T}$} & {\large $\epsilon$}\\
 &  & (ps) & (\%) \\
\hline
\hline
BC-404 & S13360-3075PE & 59 & $\geq$ 99.9\\
\hline
BC-404 & S13360-3050PE & 60 & $\geq$ 99.7\\
\hline
BC-404 & ASD-NUV3S-P-40 & 65 & $\geq$ 99.0\\
\hline
\end{tabular}
}
\end{table}

\section{SiPM Radiation hardness tests}
\label{subsec:Radiation}
\renewcommand\thefigure{B.\arabic{figure}}    
\renewcommand\thetable{B.\arabic{table}}  
\setcounter{figure}{0}    
\setcounter{table}{0}    

Radiation damage causes an increase of dark current and a decrease of the SiPM analog signal amplitude, resulting in an expected degradation of SiPM performance.
The MUSE BH SiPMs are positioned 5 cm away from the beam center, for SiPMs on central paddles, where the particle flux is about 500 times less than in the center of the beam---see Figure~\ref{Beam_Profile}.
As a result, the radiation dose to the SiPMs over the 9,000 hours of the experiment can be achieved in approximately 18 hours with them positioned at the center of the beam.

\begin{figure}[h]
\centering
\includegraphics[width=0.49\textwidth]{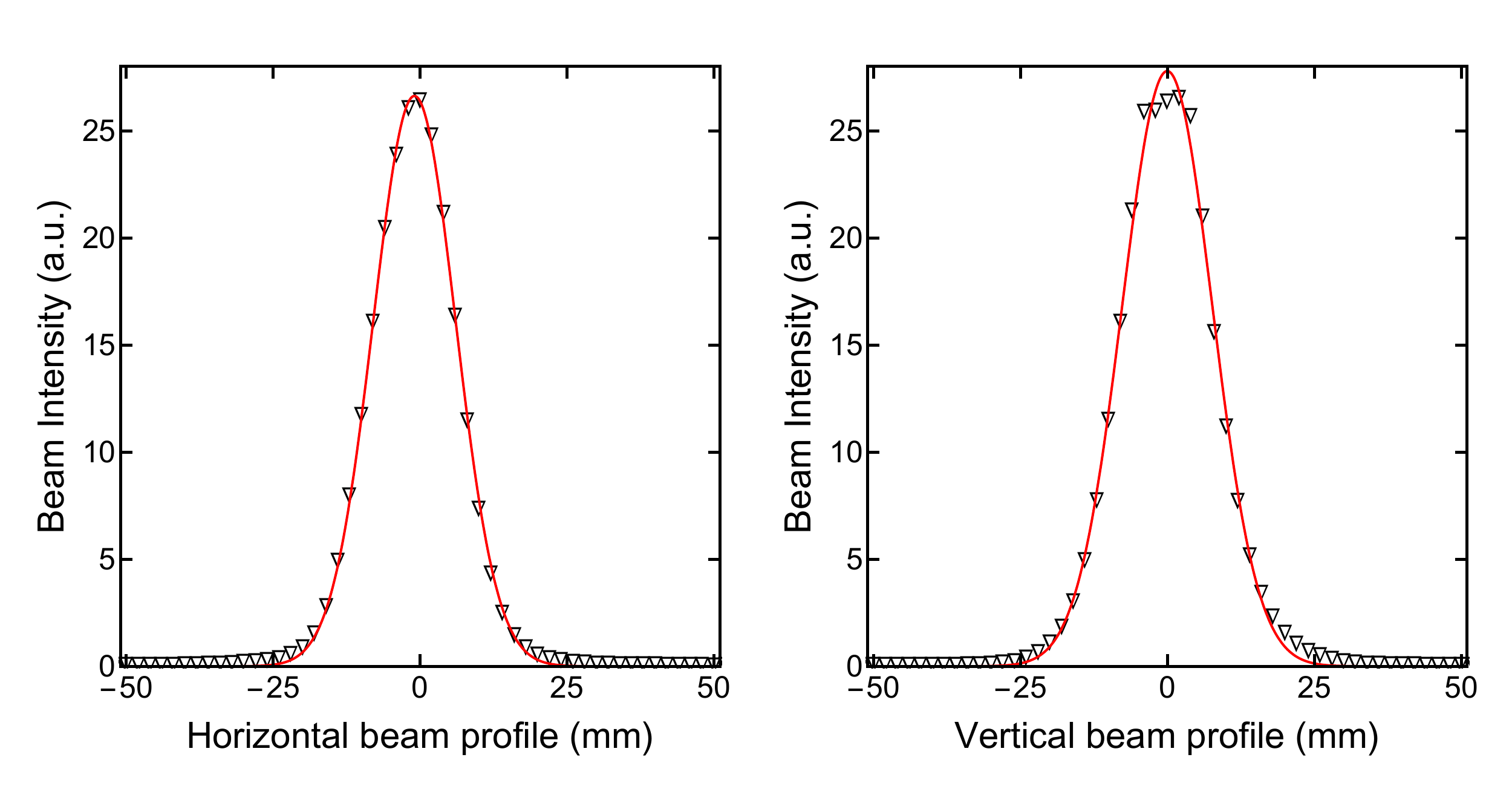}
\vskip-1mm\caption{Beam profile at the position of BH planes, measured by scanning a 2x2 mm$^2$ scintillator across the beam. Radiation is approximately 500 times less at $\pm$ 5 cm away from the center than in the center of the beam.}
\label{Beam_Profile}
\end{figure}

Three prototype detectors, with AdvanSiD ASD-NUV3S-P-40, Hamamatsu S13360-3050PE and Hamamatsu S13360-3075PE SiPMs glued at both ends of the scintillators were checked for the performance degradation after radiation damage.
The SiPMs were put into the beam line center one by one and were irradiated.
A reference 2x2x2 mm$^3$ plastic-scintillator detector was installed on the back of SiPMs for rate measurements.
This detector, at the center of the beam spot, intercepts roughly 1.1\% of the beam, corresponding to a
38.5 kHz rate at the planned 3.5 MHz MUSE beam flux. Data were taken at 161 MeV/$c$ with positive polarity. 
The beam composition was about 53\% $\pi$'s, 40\% $e$'s, and 7\% $\mu$'s.

The AdvanSiD SiPMs were irradiated in 1 hour steps, up to 5 hours of irradiation, with a total beam flux of roughly 1.5 MHz and approximately 16 kHz in the reference 2x2x2 mm$^3$ detector.
Due to the smaller beam flux, each hour of irradiation corresponded to 2.3\% of the expected integrated exposure during the full experiment.
The full integrated exposure was not achieved because of technical issues.
The performance of the detector was checked between irradiation times. 
The test results are shown in Table~\ref{table:Radiation_AdvanSiD}. 
The enormous increase observed in the leakage current is accompanied with a significant worsening of the time resolution and a slight drop in the efficiency.

\begin{table}[h]
\noindent\caption{AdvanSiD ASD-NUV3S-P-40 radiation hardness test results. 
For comparison with later tests, 5 h irradiation here is equivalent to 1.7 hours of irradiation with about 47 kHz on the 2x2x2 mm$^3$ detector. Data were taken at 161 MeV/$c$ with positive polarity. 
The beam composition was about 53\% $\pi$'s, 40\% $e$'s, and 7\% $\mu$'s.}
\vspace*{2mm}
\label{table:Radiation_AdvanSiD}
\centering
\noindent{\footnotesize
\begin{tabular}{|c|c|c|c|c|}
\hline
Irradiation & HV & I & {\large $\sigma$}{\small $_{T}$} & {\large $\epsilon$}\\

(\% of total) & (V) & ($\mu A$) & (ps) & (\%) \\
\hline
\hline
0 & 30 & 2.5 & 78 & 99.7\\
\hline
2.3 & 30 & 51.5 & 80 & 99.6\\
\hline
4.6 & 30 & 79 & 82 & 99.5\\
\hline
6.9 & 30 & 103 & 83 &  99.5\\
\hline
9.2 & 30 & 125 & 85 &  99.5\\
\hline
11.5 & 30 & 145 & 88 &  99.4\\
\hline
\end{tabular}
}
\end{table}

The Hamamatsu SiPMs were irradiated with a flux of roughly 47 kHz in the 2x2x2 mm$^3$ detector.
Exposures were in approximately 5 hour steps (S13360-3050PE-s to 20 hour and S13360-3075PE-s to 15 hours of total irradiation time), corresponding to 34\% of the expected total integrated exposure for each 5-hour step. 
The detector performance was checked in between the irradiation cycles.
The results of these irradiation tests are shown in Tables~\ref{table:Radiation_3050} and \ref{table:Radiation_3075}.
Again we see increases in the leakage current, worse time resolution, and (only for the 3050) a decrease in the efficiency.

\begin{table}[h]
\noindent\caption{Hamamatsu S13360-3050PE radiation hardness test results. 
For a fixed 55 V input voltage, the signal amplitude drops from 250 mV to 180 mV over 20 hours of irradiation. Data were taken at 161 MeV/$c$ with positive polarity. 
The beam composition was about 53\% $\pi$'s, 40\% $e$'s, and 7\% $\mu$'s.}
\vspace*{2mm}
\label{table:Radiation_3050}
\centering
\noindent{\footnotesize
\begin{tabular}{|c|c|c|c|c|}
\hline
Irradiation & HV & I & {\large $\sigma$}{\small $_{T}$} & {\large $\epsilon$} \\
(\% of total) & (V) & ($\mu A$) & (ps) & (\%) \\
\hline
\hline
0  & 55 & 2 & 78 & 99.7\\
\hline
34  & 55 & 83 & 100 & 98.8\\
\hline
68 & 55 & 135 & 113 & 96.4\\
\hline
109 & 55 & 190 & 118 &  97.9\\
\hline
136 & 55 & 220 & 123 &  97.1\\
\hline
\hline
2.5 months later & 55 & 150 & 113 & 94.8\\
\hline
1 day under 60$^{\circ}$C & 55 & 140 & 126 & $>$90.0\\
\hline
\end{tabular}
}
\end{table}

\begin{table}[h]
\noindent\caption{Hamamatsu S13360-3075PE radiation hardness test results. 
After 2 months, the SiPM leakage current had partly recovered.
(Resolution and efficiency were not checked.) Data were taken at 161 MeV/$c$ with positive polarity. 
The beam composition was about 53\% $\pi$'s, 40\% $e$'s, and 7\% $\mu$'s.}
\vspace*{2mm}
\label{table:Radiation_3075}
\centering
\noindent{\footnotesize
\begin{tabular}{|c|c|c|c|c|}
\hline
Irradiation & HV & I & {\large $\sigma$}{\small $_{T}$} & {\large $\epsilon$} \\
(\% of total) & (V) & ($\mu A$) & (ps) & (\%) \\
\hline
\hline
0 & 55 & 0.7 & 63 & 99.2\\
\hline
34 & 55 & 138 & 66 & 99.4\\
\hline
68 & 55 & 235 & 72 & 99.4\\
\hline
102 & 55 & 285 & 78 &  99.4\\
\hline
2 months later & 55 & 153 & - & - \\
\hline
\end{tabular}
}
\end{table}

SiPM performance tends to recover over time when they are not exposed to radiation---see Tables~\ref{table:Radiation_3050} and \ref{table:Radiation_3075}.
The S13360-3075PE SiPMs performance, tested after 2 months, shows partial recovery.
The prototype scintillator with S13360-3050PE SiPMs was put in a $60^{\circ}$ C oven for 1 day to check using heat to decrease the recovery time. 
While the SiPM recovered faster, the scintillator surface lost its reflectivity, harming the time resolution.

One concern for radiation-damaged SiPMs is that the detector is affected by significant heat from the increased current flowing through the SiPMs.
To investigate this, we glued a PT100 temperature sensor directly on an S13360-3050PE SiPM, irradiated to 136\%
of the expected total dose, with a leakage current of 220 $\mu A$.
With the SiPM sealed in its light-tight holding frame, we monitored its temperature over a 7 day operational period.
The temperature was roughly 26$^\circ$C, consistent with room temperature, with corresponding day-night temperature variations below 1$^\circ$C.
Thus, there is no indication that heat generated by the SiPMs will be an issue in the MUSE configuration.

As a result of these tests, we decided to use BC-404 scintillator paddles in combination with Hamamatsu S13360-3075PE SiPMs.

\section{Amplification Circuit}
\label{subsec:AmpCircuit}
\renewcommand\thefigure{C.\arabic{figure}}    
\renewcommand\thetable{C.\arabic{table}}  
\setcounter{figure}{0}    
\setcounter{table}{0}    

\begin{figure*}[tbh]
\centering
\includegraphics[scale=0.45, angle = 90]{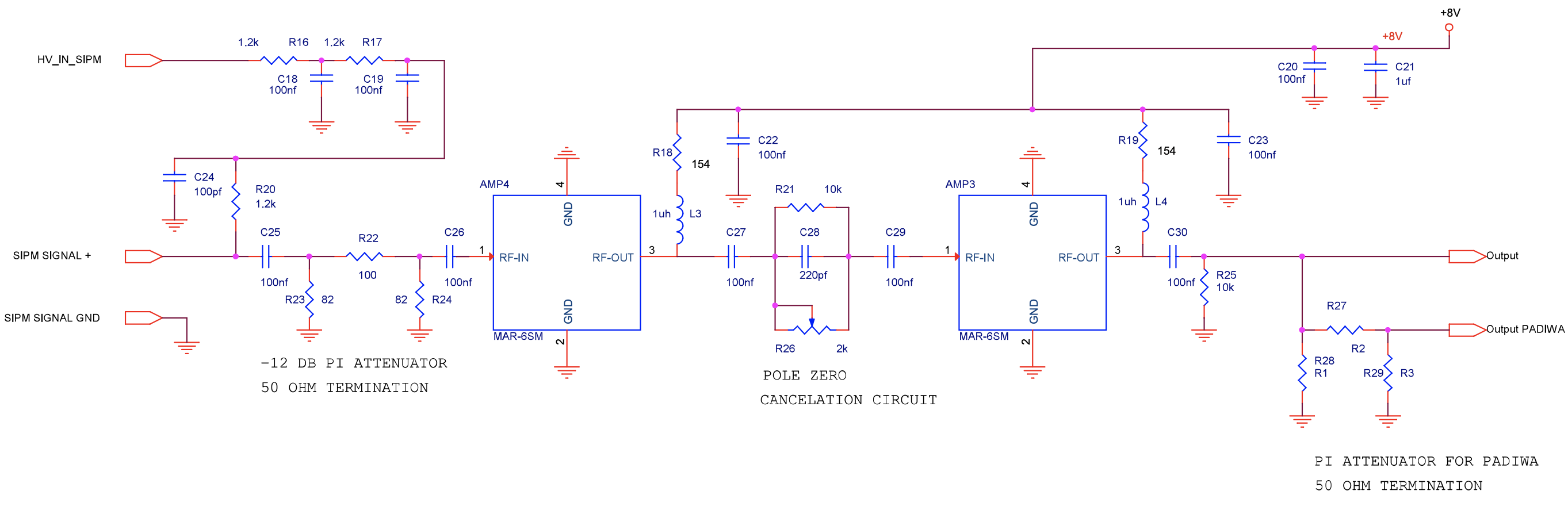}
\vskip-1mm\caption{TAU implementation of the SiPM signal amplifier circuit, based on the design of Urs Greuter (PSI).}
\label{Urs_Amplifier}
\end{figure*}

Analog signals from SiPMs in all three detectors are amplified by the same type of amplifiers. 
Figure~\ref{Urs_Amplifier} shows 
a single channel amplifier circuit adapted from 
the MAR-Amplifier design of Urs Greuter (PSI).
The amplified signal has a 1.3 (3.3) ns rise (fall) time and typically has a few hundred mV peak.

\clearpage 
\vspace*{0.25in}\hrule\vspace*{0.25in}

%\begin{multicols}{1}
%\clearpage 
%\section*{\refname}
%\addcontentsline{toc}{section}{\refname}

  %references with bibtex

%\end{multicols}
\end{twocolumn}
  \end{document}